\documentclass[notitlepage,preprintnumbers,prd,longbibliography,showpacs,nofootinbib]{revtex4-1}
\usepackage{amsmath,amssymb}
\usepackage{graphics,color}
\usepackage{caption,subcaption}
\usepackage{epsfig}
\usepackage{dcolumn}
\usepackage{multirow}
\usepackage{bm}
\usepackage{verbatim}

\usepackage[colorlinks=true,
urlcolor=blue,
linkcolor=blue,
citecolor=blue,
bookmarks=true,
bookmarksopen=true,
bookmarksnumbered=true]{hyperref}

\begin{document}
\title{$T^a_{c\bar{s}0}(2900)$, $T_{cs0}^*(2900)^0$, and other singly heavy tetraquark states}
\author{Zi-Long Man$^{1,2,3,4,5,6}$}\email{manzl@lzu.edu.cn}
\author{Yu-Nan Liu$^7$}
\author{Yan-Rui Liu$^8$}\email{yrliu@sdu.edu.cn}

\affiliation{
$^1$Department of Physics, Lanzhou University of Technology, Lanzhou 730050, China\\
$^2$School of Physical Science and Technology, Lanzhou University, Lanzhou 730000, China\\
$^3$Lanzhou Center for Theoretical Physics, Lanzhou University, Lanzhou 730000, China\\
$^4$Key Laboratory of Quantum Theory and Applications of MoE, Lanzhou University, Lanzhou 730000, China\\
$^5$Key Laboratory of Theoretical Physics of Gansu Province, Lanzhou University, Lanzhou 730000, China\\
$^6$Research Center for Hadron and CSR Physics, Lanzhou University and Institute of Modern Physics of CAS, Lanzhou 730000, China\\
$^7$ School of Science, Jilin University of Chemical Technology, Jilin, 132022, China\\
$^8$School of Physics, Shandong University, Jinan, Shandong 250100, China\\
}	

\date{\today}

\begin{abstract}
We systematically study the mass spectra of $S$-wave singly heavy tetraquark states $Qq\bar{q}\bar{q}$ ($Q=c,b$; $q=u,d,s$) in a mass splitting model. We adopt the assumption that the $X(4140)$ is the lowest $J^{PC}=1^{++}$ $cs\bar{c}\bar{s}$ tetraquark and use this state as a reference to determine the mass splittings. According to the obtained results, we also estimate the rearrangement decay widths of the tetraquarks within a simple scheme. We find that the recently observed states $T^a_{c\bar{s}0}(2900)^{++/0}$ and  $T_{cs0}^*(2900)^0$  by the LHCb Collaboration can be consistently interpreted as the second highest $I(J^P)=1(0^+)$ $cn\bar{s}\bar{n}$ ($n=u,d$) and the higher $I(J^P)=0(0^+)$ $cs\bar{n}\bar{n}$ tetraquark states, respectively. We predict several narrow tetraquark candidates: the lowest $cn\bar{s}\bar{n}$ and $cs\bar{n}\bar{n}$ with $I(J^P)=0(0^+)$ and $0(1^+)$, and their bottom counterparts. The obtained information from mass spectrum and rearrangement decay properties  will help search for the new singly heavy tetraquark states.

\end{abstract}


\maketitle
	
\section{Introduction}\label{secI}

Recently, two $J^P=0^+$ resonance states $T^a_{c\bar{s}0}(2900)^0$ and $T^a_{c\bar{s}0}(2900)^{++}$ were observed by the LHCb Collaboration in the $D_s\pi$ invariant mass distributions of the decay $B^0\to\bar{D}^0D^+_s\pi^-$ and $B^+\to D^-D^+_s\pi^+$, respectively \cite{LHCb:2022sfr,LHCb:2022lzp}. The mass and width for the former state are $M=2892\pm14\pm15$ MeV and $\Gamma=119\pm26\pm13$ MeV, respectively, and the repective data for the latter state are $M=2921\pm17\pm20$ MeV and $\Gamma=137\pm32\pm17$ MeV. Their common mass and width are determined to be $M=2908\pm11\pm20$MeV and $\Gamma=136\pm23\pm13$ MeV by assuming that they belong to the same isospin triplet. The minimal quark contents of these two states are $c\bar{s}u\bar{d}$ and $c\bar{s}\bar{u}d$, respectively, i.e., they are states with four different flavors.

Before the announcement of $T^a_{c\bar{s}}(2900)^0$ and $T^a_{c\bar{s}}(2900)^{++}$, two other exotic resonances $X_0(2900)$ $(J=0)$ and $X_1(2900)$ ($J=1$) with minimal quark content $\bar{c}\bar{s}ud$ were observed by LHCb in the $D^-K^+$ channel of the $B^+\to D^+D^-K^+$ decay \cite{LHCb:2020bls,LHCb:2020pxc}. Their masses and widths are determined to be
\begin{eqnarray*}
&&X_0(2900):\quad m=2866\pm7\pm2\;\mathrm{MeV},\quad \Gamma=57\pm12\pm4\;\mathrm{MeV};\\
&&X_1(2900):\quad m=2904\pm5\pm1\;\mathrm{MeV},\quad\Gamma=110\pm11\pm4\mathrm{MeV}.
\end{eqnarray*}
Their parities should be even and odd, respectively. According to the new naming scheme \cite{Gershon:2022xnn}, these two states are referred to as $T_{cs0}^*(2900)^0$ and $T_{cs1}^*(2900)^0$, respectively. Subsequent experimental studies have further clarified the existence and properties of them.
In 2024, both resonances were later confirmed in the $D^+K^-$ invariant-mass spectrum of the $B^- \to D^{-}D^+K^-$ decay \cite{LHCb:2024vfz}. More recently, in 2025, the $T_{cs0}^*(2900)^0$ state was further established in the $B^-\to D^-D^0 K^0_S$ decay channel \cite{LHCb:2024xyx}. In contrast, no significant evidence for the $T_{cs1}^*(2900)^0$ state was found in the $D^0K_S^0$ invariant-mass spectrum.

Besides the above states around 2.9 GeV, the D0 Collaboration had reported the evidence of a near-threshold tetraquark structure $X(5568)$ with mass $5567.8\pm2.9^{+0.9}_{-1.9}$ MeV and width $\Gamma=21.9\pm6.4^{+5.0}_{-2.5}$ MeV in the $B^0_s\pi^{\pm}$ invariant mass distribution in 2016 \cite{D0:2016mwd}. This state also has constituent quarks with four different flavors ($b,s,u,d$). Nonetheless, its existence was not confirmed by the LHCb \cite{LHCb:2016dxl}, CMS \cite{CMS:2017hfy}, CDF \cite{CDF:2017dwr}, and ATLAS \cite{ATLAS:2018udc} collaborations.

In the literature, the above exotic states were discussed in the compact tetraquark scenario \cite{Liu:2016ogz,Agaev:2016mjb,Wang:2016tsi,Wang:2016mee,Zanetti:2016wjn,Chen:2016mqt,Agaev:2016ijz,Dias:2016dme,Wang:2016wkj,Tang:2016pcf,Lu:2016zhe,Chen:2017rhl,Zhang:2017xwc,Guo:2021mja,Zhang:2020oze,Wang:2020xyc,Ozdem:2022ydv,Lu:2020qmp,Wang:2020prk,Yang:2021izl,Tan:2020cpu,Narison:2021vfl,Xue:2020vtq,An:2022vtg,Agaev:2022eyk,Yang:2023evp,Ortega:2023azl,deOliveira:2023hma,Lian:2023cgs,Liu:2022hbk,Mutuk:2020igv,Mutuk:2025hql,Mohan:2026blk},
the molecule scenario \cite{Kang:2016zmv,Chen:2016ypj,Lu:2016kxm,Sun:2016tmz,Ke:2016oez,Molina:2020hde,Hu:2020mxp,Chen:2021tad,Dong:2020rgs,Dai:2022htx,He:2020btl,Liu:2020nil,Huang:2020ptc,Wang:2021lwy,Xiao:2020ltm,Agaev:2022duz,Ke:2022ocs,Duan:2023qsg,Ding:2025uhh,Duan:2023lcj,Wang:2026prg,Song:2025ueo,Yu:2025xip,Yeo:2026dgo}, and the nonresonance scenarios such as triangle singularity \cite{Liu:2020orv,Ge:2022dsp} and kinematic effects \cite{Burns:2020epm}. According to the molecular interpretations, the  $T_{cs0}^*(2900)^0$  can be assigned as a $\bar{D}^*K^*$ bound state with $I(J^P)=0(0^+)$.  In Ref. \cite{Wang:2020prk}, the authors calculated the masses and strong decays of $\bar{c}\bar{s}qq$ states in the quark model by considering the Coulomb, liner confinement, and hyperfine interactions. Their results suggest that the $T_{cs0}^*(2900)^0$  can be interpreted as a tetraquark state with $I(J^P)=1(0^+)$. Until now, the inner structures and decay properties of  these singly heavy exotic states are still controversial.

In our previous work \cite{Cheng:2020nho}, the mass spectra and rearrangement decay widths of four singly heavy tetraquark states were investigated within the framework of the color-magnetic interaction model. The predicted masses of the $cs\bar{u}\bar{d}$ and $cu\bar{s}\bar{d}$ configurations were found to be consistent with the subsequently observed $T_{cs0}^*(2900)^0$  and  $T^a_{c\bar{s}0}(2900)^{++/0}$ states \cite{LHCb:2022sfr,LHCb:2022lzp,LHCb:2020bls,LHCb:2020pxc}. The masses were calculated by assuming that the $X(4140)$ is a $1^{++}$ $cs\bar{c}\bar{s}$ tetraquark state, and the rearrangement decay widths were obtained by assuming that the decay Hamiltonian is a constant. The results indicated that the interpretation of $X(5568)$ as a compact $bu\bar{s}\bar{d}$ tetraquark state is not supported. However, in the previous work \cite{Cheng:2020nho}, the estimated rearrangement decay widths of most tetraquarks, obtained with an inappropriate decay parameter, were a few MeV, which are smaller than the observed values.
In addition, a systematic analysis of singly heavy tetraquark states, based on their masses and rearrangement decay properties, is still missing. 
 In this article, we update our decay parameter and systematically investigate the masses and rearrangement decay properties of singly heavy tetraquark states with the configuration $Qq\bar{q}\bar{q}$ ($Q=c,b$; $q=u,d,s$).
We aim to reveal the inner  structures of  $T_{cs0}^*(2900)^0$  and $T_{c\bar{s}0}^a(2900)^{0/++}$ and to give predictions for other singly heavy tetraquark states.

This paper is arranged as follows. In Sec. \ref{secII}, we introduce our mass splitting model and a simple rearrangement scheme. In Sec. \ref{secIII}, we present the numerical results for the singly heavy tetraquark states. A brief summary is given in the last section.

\section{FORMALISM}\label{secII}

\subsection{Mass splitting model}
In this work, we calculate the mass splittings between $S$-wave compact tetraquark states with the simple chromomagnetic interaction (CMI) model in which the Hamiltonian reads
\begin{eqnarray}\label{hamiltonian}
H&=&\sum_i m_i+H_{\mathrm{CMI}}=\sum_i m_i-\sum_{i<j}C_{ij}{\lambda}_i\cdot{\lambda}_j\sigma_i\cdot\sigma_j.
\end{eqnarray}
Here, $m_i$ is effective quark mass for the $i$th constituent quark, which contains the contributions from the kinetic energy, chromoelectric interaction, and color confinement. The coupling constant $C_{ij}$ reflects the strength between the $i$th and $j$th constituent quark components. The values of $m_i$ and $C_{ij}$ can be determined from masses of conventional hadrons. $\lambda_i$ and $\sigma_i$ are the Gell-Mann and Pauli matrices for the $i$th constituent quark, respectively. One can estimate the masses of singly heavy tetraquarks with
\begin{eqnarray}\label{dia}
M&=&\sum_im_i+E_{CMI}
\end{eqnarray}
where the eigenvalue $E_{CMI}$ for a tetraquark state is obtained after the diagonalization of the CMI matrix $\langle{H_{CMI}}\rangle$.
Although the effective constituent quark masses phenomenologically absorb the average spin-independent interactions, including the chromoelectric interaction, this approximation is not exact because the chromoelectric interaction depends on the color configuration of the multiquark system and therefore cannot be completely incorporated into a universal set of effective quark masses. As found in our previous studies~\cite{Cheng:2020nho,Wu:2016gas,Wu:2018xdi,Li:2018vhp,Liu:2019zoy,Wu:2017weo,Wu:2016vtq,Li:2023wxm,Li:2023aui,Li:2025fmf,Li:2023wug}, Eq. (2) usually leads to tetraquark masses higher than the experimental values, indicating that part of the spin-independent attraction is still missing. This overestimation mainly originates from the use of a universal set of effective constituent quark masses, whose values may vary from one hadronic system to another. An alternative treatment is to introduce an explicit chromoelectric interaction, as adopted in the extended CMI models~\cite{Hogaasen:2013nca,Weng:2018mmf}. In the present work, instead of introducing additional chromoelectric parameters, we employ the reference-mass prescription to effectively compensate for the missing spin-independent contributions without introducing additional model parameters.
Here, we treat the results using Eq. \eqref{dia} as upper limits for the tetraquark masses.

In order to obtain more appropriate numerical results, we modify the mass formula Eq. \eqref{dia} to be
\begin{eqnarray}\label{ref}
	M=[M_{ref}-(E_{CMI})_{ref}]+E_{CMI}.
\end{eqnarray}
Here, $M_{ref}$ is the measured mass of the reference system and $(E_{CMI})_{ref}$ is the calculated eigenvalue of its $\langle H_{CMI}\rangle$. We assume that the needed attraction by Eq. \eqref{dia} is embodied in $M_{ref}$ which can be taken as the threshold of a hadron-hadron state or the mass of an exotic state. 

In the former case, however, there may be several choices for the threshold and the problem which choice gives more reasonable multiquark masses is difficult to  be solved. Previous studies \cite{Hyodo:2012pm,Wu:2016gas,Wu:2018xdi,Chen:2016ont,Luo:2017eub,Zhou:2018bkn,Liu:2019zoy,Li:2023aui} indicated that the estimated masses using hadron-hadron thresholds are usually lower than the experimental values. This means that the effective attractions in the CMI model differ between multiquark states and conventional hadrons. Here, we use Eq. \eqref{ref} to find lower limits for the tetraquark masses.
In the latter case,  we assign an exotic state to be a compact tetraquark and utilize  he reference state to estimate other tetraquark masses.
In the present work, we choose the $X(4140)$ as the reference state based on the following considerations. First, the $X(4140)$ has been firmly established as a $J/\psi \phi$ resonance across various experiments, with its quantum numbers  determined to be $J^{PC} = 1^{++}$ . Second, another exotic state, $X(4274)$, observed in the same decay channel  also shares the $J^{PC} = 1^{++}$ assignment \cite{CDF:2011pep,LHCb:2021uow}. Within the compact $cs\bar{c}\bar{s}$ tetraquark picture, these two states can be consistently interpreted as members of the same multiplet \cite{Wu:2016gas,Stancu:2009ka}. Furthermore, our systematic study on the reference state selection problem in Ref. \cite{Li:2023wxm} showed that taking $X(4140)$ as the benchmark state leads to a more reasonable and self-consistent description of the remaining $cs\bar{c}\bar{s}$ tetraquark candidates. For these reasons, we adopt $X(4140)$ as the reference state in the present work.
 
 In the present study, we take the latter method and express the masses of various singly heavy tetraquark states as ($\tilde{m}\equiv M_{X(4140)}-(E_{CMI})_{X(4140)}$)
\begin{eqnarray}\label{mass}
	M_{cn\bar{n}\bar{n}}&=&\tilde{m}+E_{CMI}-2\Delta_{sn}-\Delta_{cn},\label{formula-cnnn}\\
	M_{cn\bar{s}\bar{n}}&=&\tilde{m}+E_{CMI}-\Delta_{sn}-\Delta_{cn},\label{formula-cnsn}\\
	M_{cn\bar{s}\bar{s}}&=&\tilde{m}+E_{CMI}-\Delta_{cn},\label{formula-cnss}\\
	M_{cs\bar{n}\bar{n}}&=&\tilde{m}+E_{CMI}-\Delta_{sn}-\Delta_{cn},\label{formula-csnn}\\
	M_{cs\bar{s}\bar{n}}&=&\tilde{m}+E_{CMI}-\Delta_{cn},\label{formula-cssn}\\
	M_{cs\bar{s}\bar{s}}&=&\tilde{m}+E_{CMI}-\Delta_{cs},\label{formula-csss}\\
	M_{bn\bar{n}\bar{n}}&=&\tilde{m}+E_{CMI}-2\Delta_{sn}-\Delta_{cn}+\Delta_{bc},\label{formula-bnnn}\\
	M_{bn\bar{s}\bar{n}}&=&\tilde{m}+E_{CMI}-\Delta_{sn}-\Delta_{cn}+\Delta_{bc},\label{formula-bnsn}\\
	M_{bn\bar{s}\bar{s}}&=&\tilde{m}+E_{CMI}-\Delta_{cn}+\Delta_{bc},\label{formula-bnss}\\
	M_{bs\bar{n}\bar{n}}&=&\tilde{m}+E_{CMI}-\Delta_{sn}-\Delta_{cn}+\Delta_{bc},\label{formula-bsnn}\\
	M_{bs\bar{s}\bar{n}}&=&\tilde{m}+E_{CMI}-\Delta_{cn}+\Delta_{bc},\label{formula-bssn}\\
	M_{bs\bar{s}\bar{s}}&=&\tilde{m}+E_{CMI}-\Delta_{cs}+\Delta_{bc},\label{formula-bsss}
\end{eqnarray}
where we introduce effective quark mass gaps $\Delta_{sn}=m_s-m_n$, $\Delta_{cs}=m_c-m_s$, $\Delta_{cn}=m_c-m_n$, and $\Delta_{bc}=m_b-m_c$ with $n=u,d$. These four parameters can be determined from various conventional hadrons.

\begin{table}[htbp]
\caption{Spin and color wave functions (w.f.) for singly heavy tetraquark states in the diquark-antidiquark base. In the spin wave functions, a superscript represents the total spin of diquark/antidiquark or tetraquark. In the color wave functions, the subscript $3_c$, $6_c$, $1_c$, $\bar{3}_c$, or $\bar{6}_c$ denotes the color representation.}\label{wave function 1}
	\centering
	\begin{tabular}{c|ccc}\hline\hline
		\multirow{6}*{\text{Spin w.f.}}
		&$\chi_1=[(Q_1q_2)^1(\bar{q}_3\bar{q}_4)^1]^{J=2}$\\
		&$\chi_2=[(Q_1q_2)^1(\bar{q}_3\bar{q}_4)^1]^{J=1}$\\
		&$\chi_4=[(Q_1q_2)^1(\bar{q}_3\bar{q}_4)^0]^{J=1}$\\
		&$\chi_5=[(Q_1q_2)^0(\bar{q}_3\bar{q}_4)^1]^{J=1}$\\
		&$\chi_3=[(Q_1q_2)^1(\bar{q}_3\bar{q}_4)^1]^{J=0}$\\
		&$\chi_6=[(Q_1q_2)^0(\bar{q}_3\bar{q}_4)^0]^{J=0}$\\
		\hline
		\text{Color w.f.}&$\phi_1=[(Q_1q_2)_{6_c}(\bar{q}_3\bar{q}_4)_{\bar{6}_c}]_{1_c}$\\
		&$\phi_2=[(Q_1q_2)_{\bar{3}_c}(\bar{q}_3\bar{q}_4)_{3_c}]_{1_c}$\\\hline\hline
	\end{tabular}
\end{table}
\begin{table}[htbp]
\caption{The twelve possible $color\otimes spin$ wave function bases for the $S$-wave $Q_{1}q_{2}\bar{q}_{3}\bar{q}_{4}$ tetraquark states. They are given with the notation $|(Q_{1}q_{2})_{color}^{spin}(\bar{q}_{3}\bar{q}_{4})_{color}^{spin}\rangle^{spin}$. The symbol $\delta_{34}^{S,A}$ reflects the constraint from the Pauli principle. If the flavor wave function for the two light antiquarks is symmetric (antisymmetric), one has $\delta_{34}^S=0$ ( $\delta_{34}^A=0$). When the factor cannot be 0, its value is 1.} \label{color-spin wave}
	\centering
	\begin{tabular}{c|l|ll}\hline\hline
		$J=2$&$\phi_1\chi_1=|(Q_{1}q_{2})_{6_c}^{1}(\bar{q}_{3}\bar{q}_{4})_{\bar{6}_c}^{1}\rangle^{2}_{1_c}\delta_{34}^S$&$\phi_2\chi_1=|(Q_{1}q_{2})_{\bar{3}_c}^{1}(\bar{q}_{3}\bar{q}_{4})_{3_c}^{1}\rangle^{2}_{1_c}\delta_{34}^A$\\
		\hline
		$\multirow{3}*{J=1}$&$\phi_1\chi_2=|(Q_{1}q_{2})_{6_c}^{1}(\bar{q}_{3}\bar{q}_{4})_{\bar{6}_c}^{1}\rangle^{1}_{1_c}\delta_{34}^S$&$
		\phi_2\chi_2=|(Q_{1}q_{2})_{\bar{3}_c}^{1}(\bar{q}_{3}\bar{q}_{4})_{3_c}^{1}\rangle^{1}_{1_c}\delta_{34}^A$\\
		&$\phi_1\chi_4=|(Q_{1}q_{2})_{6_c}^{1}(\bar{q}_{3}\bar{q}_{4})_{\bar{6}_c}^{0}\rangle^{1}_{1_c}\delta_{34}^A$&$\phi_2\chi_4=|(Q_{1}q_{2})_{\bar{3}_c}^{1}(\bar{q}_{3_c}\bar{q}_{4})_{3}^{0}\rangle^{1}_{1_c}\delta_{34}^S$ \\
		&$\phi_1\chi_5=|(Q_{1}q_{2})_{6_c}^{0}(\bar{q}_{3}\bar{q}_{4})_{\bar{6}_c}^{1}\rangle^{1}_{1_c}\delta_{34}^S$&$\phi_2\chi_5=|(Q_{1}q_{2})_{\bar{3}_c}^{0}(\bar{q}_{3_c}\bar{q}_{4})_{3}^{1}\rangle^{1}_{1_c}\delta_{34}^A$ \\
		\hline
		$\multirow{2}*{J=0}$&$\phi_1\chi_3=|(Q_{1}q_{2})_{6_c}^{1}(\bar{q}_{3}\bar{q}_{4})_{\bar{6}_c}^{1}\rangle^{0}_{1_c}\delta_{34}^S$&$\phi_2\chi_3=|(Q_{1}q_{2})_{\bar{3}_c}^{1}(\bar{q}_{3}\bar{q}_{4})_{3_c}^{1}\rangle^{0}_{1_c}\delta_{34}^A $\\
		&$\phi_1\chi_6=|(Q_{1}q_{2})_{6_c}^{0}(\bar{q}_{3}\bar{q}_{4})_{\bar{6}_c}^{0}\rangle^{0}_{1_c}\delta_{34}^A$&$\phi_2\chi_6=|(Q_{1}q_{2})_{\bar{3}_c}^{0}(\bar{q}_{3}\bar{q}_{4})_{3_c}^{0}\rangle^{0}_{1_c}\delta_{34}^S$\\\hline\hline
	\end{tabular}
\end{table}

To estimate the masses of singly heavy tetraquark states, one constructs their flavor-spin-color wave functions and then solves the eigenvalue problem for the CMI matrices. In the present work, the spin and color wave functions are constructed using the diquark-antidiquark base $[(Q_1q_2)(\bar{q}_3\bar{q}_4)]$. We list them in Table \ref{wave function 1}. Considering the constraint from the Pauli principle, one obtains the color-spin wave function bases given in Table \ref{color-spin wave}. Once the Pauli principle applies, the results are base-independent. We display the explicit CMI matrices and corresponding wave function bases for the singly heavy tetraquark states in Table \ref{CMI matrices}. To simplify the expressions, we have defined several variables: $\tau=C_{12}+C_{34}$, $\theta=C_{12}-C_{34}$, $\alpha=C_{13}+C_{14}+C_{23}+C_{24}$, $\beta=C_{13}-C_{14}+C_{23}-C_{24}$, $\mu=C_{13}+C_{14}-C_{23}-C_{24}$, $\nu=C_{13}-C_{14}-C_{23}+C_{24}$, $\gamma=3C_{12}-C_{34}$, and $\eta=C_{12}-3C_{34}$.

\setlength{\tabcolsep}{1.5mm}\begin{table}[htbp]
	\caption{The CMI matrices and their corresponding wave function (w.f.) bases for the $Q_{1}q_{2}\bar{q}_{3}\bar{q}_{4}$ tetraquark states. We use the notation $[Qn(\bar{n}\bar{n})^{I_{\bar{n}\bar{n}}}]^I$ to denote the $Qn\bar{n}\bar{n}$ states and $[Qq\bar{q}\bar{q}]^I$ to denote other states. }\scriptsize\label{CMI matrices}
	\begin{tabular}{cccc}\hline\hline
States & $J^{P}$ &$\langle H_{CMI} \rangle$& w.f. bases \\\hline
$[Qn(\bar{n}\bar{n})^1]^{\frac32,\frac12}$, $[Qs\bar{n}\bar{n}]^{1}$, $[Qn\bar{s}\bar{s}]^{\frac12}$, $[Qs\bar{s}\bar{s}]^{0}$            &$2^+$&$\frac{4}{3}(2\tau+\alpha)$&$(\phi_2\chi_1)$\\ 
		&$1^+$&$\begin{pmatrix}
			\frac{4}{3}(2\tau-\alpha) &-4\mu &-\frac{4\sqrt{2}}{3}\mu \\
			&-\frac{4}{3}\eta&-2\sqrt{2}\tau\\
			&                &-\frac{8}{3}\gamma\\
		\end{pmatrix}$&$\begin{pmatrix}\phi_2\chi_2\\\phi_1\chi_4\\ \phi_2\chi_5\end{pmatrix}$
\\
  &$0^+$&$\begin{pmatrix}\frac{8}{3}(\tau-\alpha) &2\sqrt{6}\alpha\\
			&4\tau\\\end{pmatrix}$&$\begin{pmatrix}\phi_2\chi_3\\ \phi_1\chi_6\end{pmatrix}$
\\
$[Qn(\bar{n}\bar{n})^0]^{\frac12}$, $[Qs\bar{n}\bar{n}]^{0}$             &$2^+$&$\frac{2}{3}(-2\tau+5\alpha)$&$(\phi_1\chi_1)$\\
&$1^+$&$\begin{pmatrix}
			-\frac{2}{3}(\tau+5\alpha) &-4\mu &-\frac{10\sqrt{2}}{3}\mu \\
			&\frac{8}{3}\eta&-2\sqrt{2}\alpha\\
			&                &\frac{4}{3}\gamma\\
		\end{pmatrix}$&$\begin{pmatrix}\phi_1\chi_2\\ \phi_2\chi_4 \\ \phi_1\chi_5\end{pmatrix}$\\
		&$0^+$&$\begin{pmatrix}
			-\frac{4}{3}(\tau+5\alpha) &2\sqrt{6}\alpha\\
			&-8\tau\\
		\end{pmatrix}$&$\begin{pmatrix}\phi_1\chi_3\\ \phi_2\chi_6\end{pmatrix}$\\
$[Qn\bar{s}\bar{n}]^{1,0}$, $[Qs\bar{s}\bar{n}]^{\frac12}$                      &$2^+$&$\begin{pmatrix}
			\frac{2}{3}(-2\tau+5\alpha) &2\sqrt{2}\nu\\
			&\frac{4}{3}(2\tau+\alpha)\\
		\end{pmatrix}$&$\begin{pmatrix}\phi_1\chi_1\\\phi_2\chi_1\end{pmatrix}$\\
		&$1^+$&$\begin{pmatrix}
			-\frac{2}{3}(2\tau+5\alpha) &2\sqrt{2}\nu &\frac{10\sqrt{2}}{3}\beta &-4\mu &-\frac{10\sqrt{2}}{3}\mu &4\beta \\
			&\frac{4}{3}(2\tau-\alpha)&-4\mu &\frac{4\sqrt{2}}{3}\beta &4\beta &-\frac{4\sqrt{2}}{3}\mu \\
			&                &-\frac{4}{3}\eta &0  &\frac{10}{3}\nu &-2\sqrt{2}\alpha\\
			&                 &                &\frac{8}{3}\eta &-2\sqrt{2}\alpha &\frac{4}{3}\nu\\
			&                 &                &                 &\frac{4}{3}\gamma&0                     \\
			&                  &               &                 &           &-\frac{8}{3}\gamma\\
		\end{pmatrix}$&$\begin{pmatrix}\phi_1\chi_2\\\phi_2\chi_2\\\phi_1\chi_4\\
			\phi_2\chi_4\\\phi_1\chi_5\\\phi_2\chi_5\end{pmatrix}$\\
		&$0^+$&$\begin{pmatrix}
			-\frac{4}{3}(\tau+5\alpha) &4\sqrt{2}\nu &-\frac{10}{\sqrt{3}}\nu &2\sqrt{6}\alpha\\
			&\frac{8}{3}(\tau-\alpha)&2\sqrt{6}\alpha &-\frac{4}{\sqrt{3}}\nu\\
			&2\sqrt{6}\alpha &4\tau &0\\
			&&&-8\tau\\
		\end{pmatrix}$&$\begin{pmatrix}\phi_1\chi_3\\\phi_2\chi_3\\
			\phi_1\chi_6\\\phi_2\chi_6\end{pmatrix}$\\\hline\hline		
	\end{tabular}
\end{table}

\subsection{Rearrangement decay }

The experience in the pentaquark case \cite{Cheng:2019obk,Li:2023aui} suggests that the combined analysis of spectrum and rearrangement decay properties  may restrict the assignment of quantum numbers and help us to understand the structures of observed hadrons. A similar approach can be extended to tetraquark systems, where the inclusion of decay information is likewise essential for understanding their configurations  \cite{Cheng:2020nho,Li:2023aui,Li:2023wxm}. Since the dominant two-body decay for a multiquark is through the rearrangement mechanism, in the present work, we still use the simple scheme to study the decay properties of the singly heavy tetraquark states, where the decay Hamiltonian is assumed to be a constant $H_{decay}={\mathcal C}$. It means that the short-range rescattering may be the driving force for quark rearranging into final meson-meson states. In principle, the gluon exchange and quark pair creation also contribute to the decay Hamiltonian, but we do not consider such effects in the present simple analysis.  

For the singly heavy tetraquark states, their two possible decay types are
 \begin{eqnarray}
 	\begin{array}{l}
 		(Q_1q_2)_{1c}(\bar{q_3}\bar{q_4})_{1c}\to(Q_1\bar{q_3})_{1c}+(q_2\bar{q_4})_{1c},\\
 		(Q_1q_2)_{1c}(\bar{q_3}\bar{q_4})_{1c}\to(Q_1\bar{q_4})_{1c}+(q_2\bar{q_3})_{1c}.
 	\end{array}
\end{eqnarray}	
The amplitude squared in a decay channel can be expressed as
\begin{eqnarray}
|{\cal M}|^2&=&{\mathcal C}^2|\sum_{i}x_iy_i|^2,\label{amplitude}
\end{eqnarray}	
where $x_i$'s ($y_i$'s) are coefficients of the initial (final) wave function in the diquark-antidiquark base. The values of $x_i$'s can be obtained from the eigenvector of the corresponding CMI matrix while those of $y_i$'s are calculated by recoupling the meson-meson state into the diquark-antidiquark base wave functions. Here, the constant $\mathcal{C}$ denotes the effective coupling strength that governs the transition from the compact tetraquark state to the two meson final states\cite{Cheng:2020nho,Li:2023aui,Li:2023wxm}. The matrix element for this two-body rearrangement decay is evaluated using $\mathcal{C}$, where  the flavor-spin-color transition is rigorously calculated through wave function overlaps.

Then the rearrangement decay width for a singly heavy tetraquark is
\begin{eqnarray}
	\Gamma&=&|{\cal M}|^2\frac{\mathbf{|P|}}{8\pi M^2_{Qq\bar{q}\bar{q}}},\label{decay}
\end{eqnarray}
where $M_{Qq\bar{q}\bar{q}}$ represents the tetraquark mass and ${\mathbf P}$ is the 3-momentum of a final meson in the rest frame of the initial state.
It should be noted that for several kinematically forbidden channels, although their physical partial decay widths are naturally zero due to the vanishing phase space, their dimensionless coupling matrix element $100|{\cal M}|^2/ {\mathcal C}^2$  are still presented in the tables to illustrate wave function overlaps or the coupling strengths between the tetraquark and the two-meson states.

\section{NUMERICAL RESULTS} \label{secIII}

\subsection{Model parameters}

\begin{table}[htbp]
	\caption{Effective coupling parameters $C_{ij}$'s in units of MeV.}\label{effectiveparameters}
	\centering
	\begin{tabular}{ccccccccccccccccc}
		\hline\hline
		$C_{ij}$    &$n$   &$s$   &$c$   &$b$   &$C_{i\bar{j}}$&$\bar{n}$&$\bar{s}$&$\bar{c}$&$\bar{b}$\\ \hline
		n           &18.3&12.1&4.0 &1.3 &n             &29.8&18.7&6.6 &2.1 \\
		s           &    &6.5 &4.3 &1.3 &s             &    &9.8&6.7 &2.3 \\
		c           &    &    &3.5 &2.0 &c             &    &    &5.3 &3.3 \\
		b           &    &    &    &1.9 &b             &    &    &    &2.9 \\ \hline\hline
	\end{tabular}\\
\end{table}

For the input mass of the $X(4140)$, we use $M_{X(4140)}=4146.5$ MeV \cite{Li:2023wxm}. Other parameters in the above formalism can be extracted from various hadron ground  states. We present the relevant coupling parameters $C_{ij}$'s \cite{Wu:2018xdi} in Table \ref{effectiveparameters}. The coupling parameters $C_i$ in this table have been refitted and updated by adopting the latest  masses of conventional heavy mesons and baryons from the recent review of particle physics (PDG), which guarantees the reliability  of our model inputs. The effective quark mass gaps $\Delta_{bc}=3340.2$ MeV, $\Delta_{cn}=1280.7$ MeV, and $\Delta_{sn}=90.6$ MeV have been fixed in Refs. \cite{Cheng:2020nho,Wu:2018xdi}. The following discussions also need the determination of $\Delta_{cs}$. From  results shown in Table \ref{QMD}, we choose to use the larger value $\Delta_{cs}= 1180.6 $ MeV so that more reasonable tetraquark masses can be obtained \cite{Wu:2018xdi}. Here, we also fix $\Delta_{bs}= 4520.2$ MeV in a similar way, although it is not adopted in the present study. These determined quark mass gaps satisfy approximately the relations $\Delta_{cn}\approx\Delta_{cs}+\Delta_{sn}$ and $\Delta_{bs}\approx\Delta_{bc}+\Delta_{cs}$. To estimate the upper limits for the masses of tetraquark states, we adopt $m_n=361.8$ MeV, $m_s=542.4$ MeV, $m_c=1724.1$ MeV, and $m_b=5054.4$ MeV \cite{Wu:2016gas,Wu:2018xdi,Li:2023wxm,Li:2023wug}.

\begin{table}[htbp]
	\caption{Quark mass gaps $\Delta_{cs}$ and $\Delta_{bs}$ (units: MeV) determined from masses of various conventional hadrons.}\label{QMD}
	\centering
	\begin{tabular}{ccc|ccc}\hline\hline
		Hadron&Hadron&$\Delta_{cs}$&Hadron&Hadron&$\Delta_{bs}$\\\hline
		$J/\psi$&$\phi$&1049.4&$\Upsilon$&$\phi$&4237.5\\
		$J/\psi(\eta_c)$&$D_{s}^*(D_s)$&992.2 (993.2)&$\Upsilon(\eta_b)$&$B_{s}^*(B_s)$&4041.7 (4041.8)\\
		$D^*(D)$&$K^*(K)$&1180.6 (1179.4)&$B^*(B)$&$K^*(K)$&4520.2 (4518.8)\\
		$D_s$&$\phi$&1106.6&$B_s$&$\phi$&4433.8\\
		$B_c$&$B_s$&924.1&$B_c$&$D_s$&4252.2\\
		$\Lambda_c$&$\Lambda$&1170.8&$\Lambda_b$&$\Lambda$&4503.8\\
		$\Sigma_c^*(\Sigma_c)$&$\Sigma^*(\Sigma)$&1176.2 (1178.4)&$\Sigma_b^*(\Sigma_b)$&$\Sigma^*(\Sigma)$&4506.1 (4509.5)\\
		$\Xi_c^*(\Xi_c^\prime)$&$\Xi^*(\Xi)$&1137.3 (1159.1)&$\Xi_b^*(\Xi_b^\prime)$&$\Xi^*(\Xi)$&4463.2 (4483.7)\\
		$\Omega_c^*$&$\Omega$&1100.3&$\Omega_b$&$\Omega$&4415.5\\
		$\Xi_{cc}$&$\Xi$&1112.2\\
		\hline\hline
	\end{tabular}
\end{table}

Each system has its own decay parameter ${\mathcal C}$, which can be extracted from measured widths of assigned tetraquark states. The extraction relies on an assumption that the sum of two-body partial widths for the rearrangement decay channels is equal to the measured width, $\Gamma_{sum}\approx\Gamma_{exp}$. In the $cs\bar{c}\bar{s}$ and $cc\bar{n}\bar{n}$ cases, the value around 7.3 GeV was adopted. In the present case, we employ ${\mathcal C}=13.577$ GeV. It is extracted from the measured width of $T_{c\bar{s}0}^a(2900)^{++}$ by assigning it as the second highest $I(J^P)=0(0^+)$ $cn\bar{s}n$ tetraquark state (see Sec. \ref{sec-Tcsbar}). 

The singly bottom tetraquark states are the analogs of the singly charmed ones, with the $c$ quark replaced by the $b$ quark. These two cases have similar properties in spectrum and decay. Until now, there is no experimental confirmation of a singly bottom tetraquark state, and the extraction of ${\cal C}$ is therefore not available. We assume that the bottom case has the same value of ${\mathcal C}$ as the charmed case in the following discussions. Although this treatment is crude, it is relatively reliable for studying the dominant decay modes of these states.  

The final states in tetraquark decays include conventional mesons containing an $s\bar{s}$  component. Within the quark model, the vector meson $\phi$ is well approximated as a pure  $s\bar{s}$ configuration. However, the pseudoscalar mesons $\eta$ and $\eta^{\prime}$ cannot be described by a single quark composition. Instead, they arise from the mixing of the singlet $\eta_1$ and octet $\eta_8$ states. Their wave functions can be written as
\begin{equation}
\begin{aligned}
|\eta\rangle = \cos(\theta)\,|\eta_8\rangle - \sin(\theta)\,|\eta_1\rangle, \\
|\eta'\rangle = \sin(\theta)\,|\eta_8\rangle + \cos(\theta)\,|\eta_1\rangle.
\end{aligned}
\end{equation}
where $\theta$ denotes the $\eta-\eta^{\prime}$ mixing angle.  In the present calculation, the value $\theta=-11.3^\circ$
 is adopted \cite{ParticleDataGroup:2024cfk}.

We now calculate the values of masses and rearrangement decay widths using the above parameters. The estimated results are presented in Tables \ref{spectrum-Qnnn}-\ref{decay-Qsss}. These results show that the widths for the singly bottom tetraquark states are smaller than those for the corresponding singly charm ones. The reason is that the phase space becomes smaller with increasing heavy quark mass. For clarity, we will employ the notation $\tilde{T}$ to represent the predicted states, distinguishing them from the experimental candidates. The superscript $a$ ($f$) of $T$ and $\tilde{T}$ means the isospin $I=1$ ($I=0$).

\begin{figure*}[htbp]\centering
\resizebox{0.95\textwidth}{!}{
	\begin{tabular}{ccccc}	\includegraphics[width=150pt]{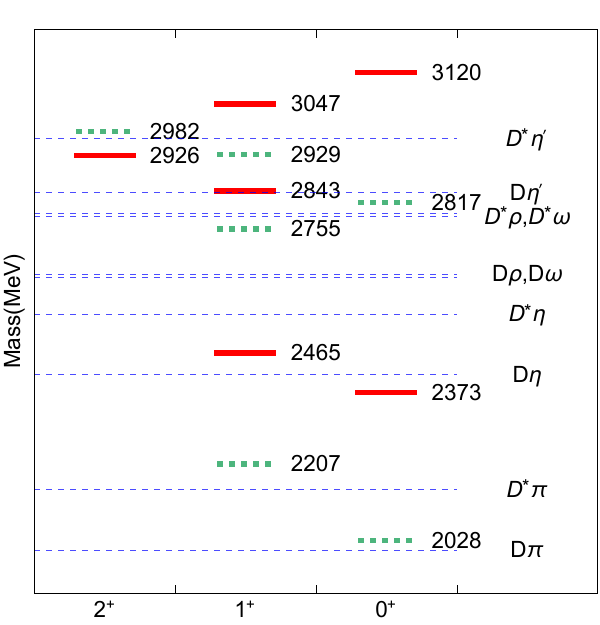}&$\qquad$&\includegraphics[width=150pt]{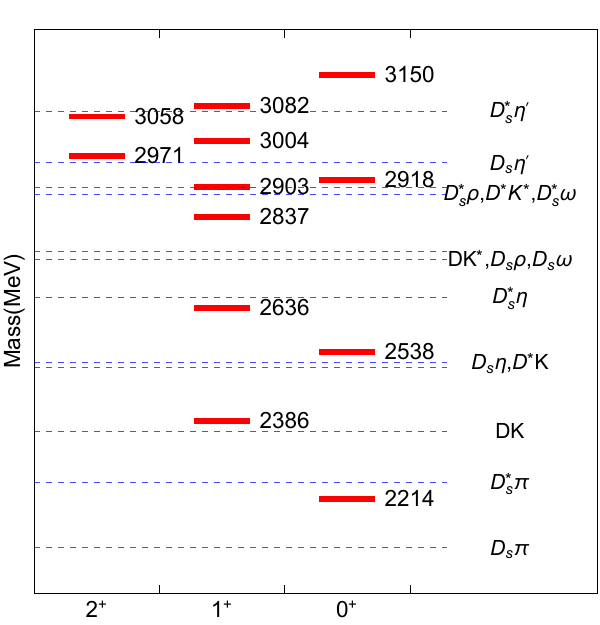}&$\qquad$&\includegraphics[width=150pt]{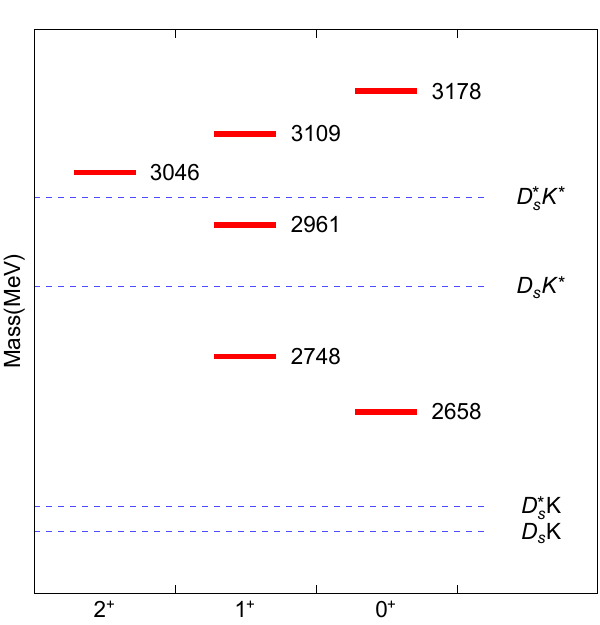}\\
		(a) $I=\frac12/\frac32$ $cn\bar{n}\bar{n}$ states & $\qquad$& (b)  $cn\bar{s}\bar{n}$ states&$\qquad$& (c)  $cn\bar{s}\bar{s}$ states\\
		\includegraphics[width=150pt]{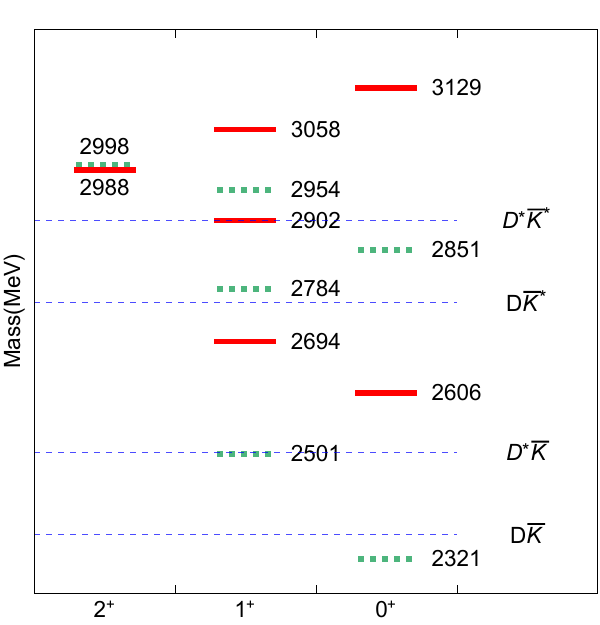}&$\qquad$&\includegraphics[width=150pt]{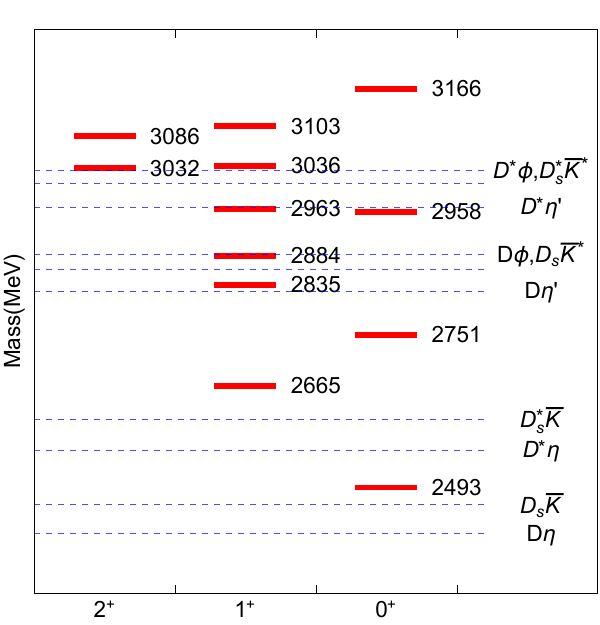}&$\qquad$&\includegraphics[width=150pt]{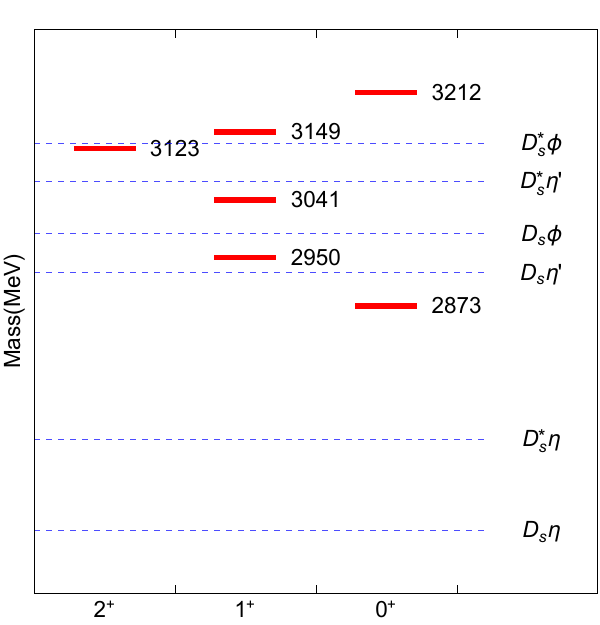}\\
		(d) $I=0/1$ $cs\bar{n}\bar{n}$ states & $\qquad$& (e)  $cs\bar{s}\bar{n}$ states&$\qquad$& (f)  $cs\bar{s}\bar{s}$ states\\
		\includegraphics[width=150pt]{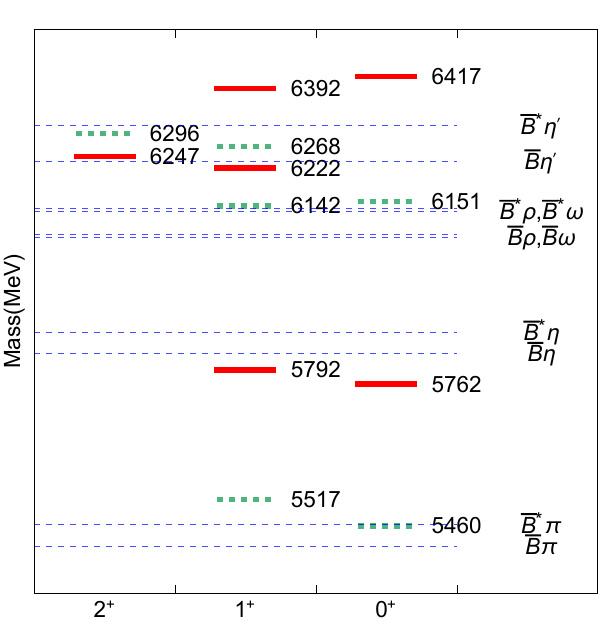}&$\qquad$&\includegraphics[width=150pt]{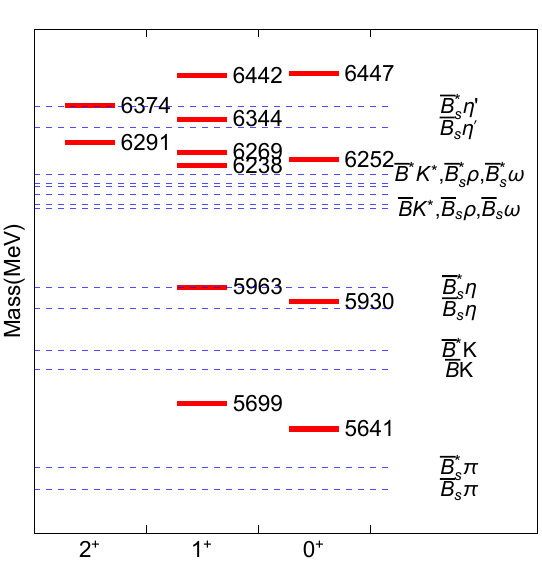}&$\qquad$&\includegraphics[width=150pt]{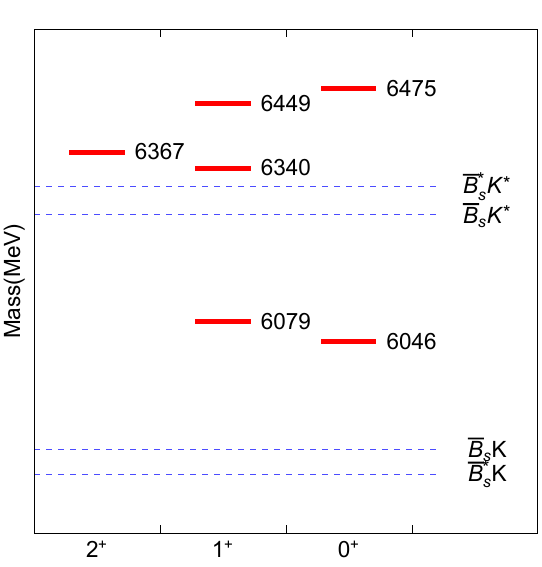}\\
		(g) $I=\frac12/\frac32$ $bn\bar{n}\bar{n}$ states & $\qquad$& (h)  $bn\bar{s}\bar{n}$ states&$\qquad$& (i) $bn\bar{s}\bar{s}$ states\\
		\includegraphics[width=150pt]{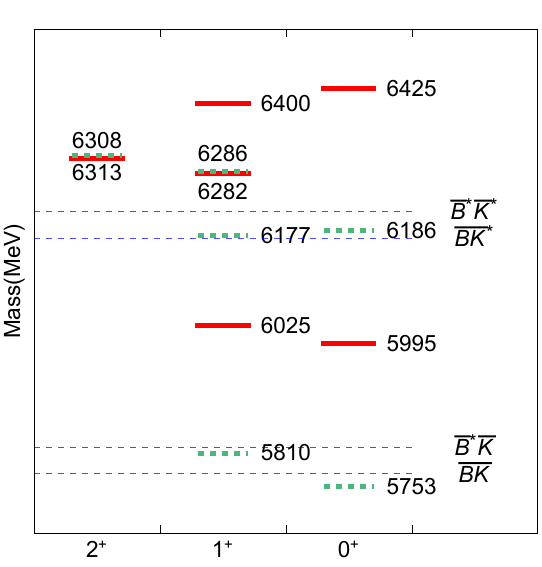}&$\qquad$&\includegraphics[width=150pt]{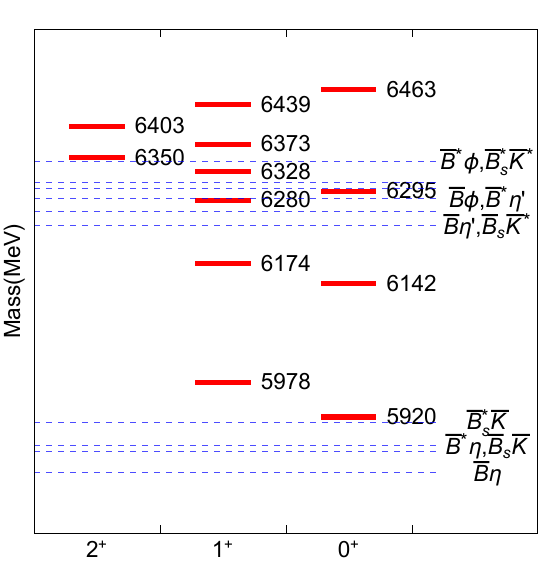}&$\qquad$&\includegraphics[width=150pt]{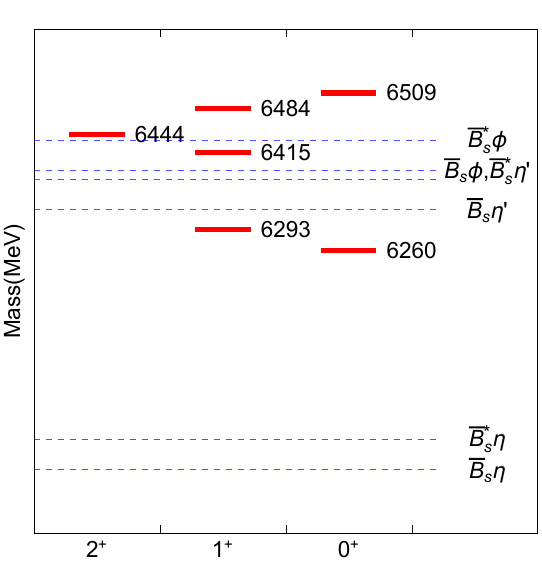}\\
		(j) $I=0/1$ $bs\bar{n}\bar{n}$ states & $\qquad$& (k)  $bs\bar{s}\bar{n}$ states&$\qquad$& (l)  $bs\bar{s}\bar{s}$ states
	\end{tabular}
	}
	\caption{Relative positions for the singly heavy tetraquark states. The red solid and green dashed lines in the panels (a), (d), (g), and (j) correspond to the tetraquark masses of the $I_{\bar{n}\bar{n}}=1$ and $I_{\bar{n}\bar{n}}=0$ states, respectively.  The blue dashed lines represent all possible thresholds. Note that the red solid lines in the panels (a), (b), (g), and (h) indicate both the highest isospin ($I_h$) states and their degenerate $(I_h-1)$ states.}\label{relative}
\end{figure*}

\subsection{The $cn\bar{n}\bar{n}$ and $bn\bar{n}\bar{n}$ systems}

\begin{table}[htbp]
\caption{Mass spectra for the $cn\bar{n}\bar{n}$ and $bn\bar{n}\bar{n}$ states in units of MeV. The thresholds of $D\pi$ and $\bar{B}\pi$ are used to establish the lower limits for the tetraquark masses in the former and latter cases, respectively.  The upper limits are obtained using the effective quark masses: $m_n=361.8$ MeV, $m_c=1724.1$ MeV, and $m_b=5054.4$ MeV. We adopt values from the mass splitting model (fourth column) in the discussions. The systems have degenerate states with different isospins.}\label{spectrum-Qnnn}\scriptsize
	\begin{tabular}{c|ccccccc}\hline
		\hline\multicolumn{6}{c}{$[cn(\bar{n}\bar{n})^{I_{\bar{n}\bar{n}}}]^{I}$ states} \\\hline
		$[I_{\bar{n}\bar{n}},I](J^{P})$ & $\langle H_{CMI} \rangle$ &Eigenvalue &Mass&Lower limits&Upper limits\\\hline
		$[1,\frac32/\frac12](2^{+})$ &$\left(\begin{array}{c}156.5\end{array}\right)$&$\left(\begin{array}{c}156.5\end{array}\right)$&$\left(\begin{array}{c}2925.7\end{array}\right)$&$\left(\begin{array}{c}2741.8\end{array}\right)$&$\left(\begin{array}{c}2966.0\end{array}\right)$\\
		$[1,\frac32/\frac12](1^{+})$ &$\left(\begin{array}{ccc}-37.6&185.6&87.5\\185.6&67.9&-205.9\\87.5&-205.9&16.8\end{array}\right)$&$\left(\begin{array}{c}277.8\\73.7\\-304.5\end{array}\right)$&$\left(\begin{array}{c}3047.0\\2842.9\\2464.7\end{array}\right)$&$\left(\begin{array}{c}2863.1\\2658.9\\2280.8\end{array}\right)$&$\left(\begin{array}{c}3087.3\\2883.2\\2505.0\end{array}\right)$\\
		$[1,\frac32/\frac12](0^{+})$ &$\left(\begin{array}{cc}-134.7&356.6\\356.6&89.2\end{array}\right)$&$\left(\begin{array}{c}351.1\\-396.5\end{array}\right)$&$\left(\begin{array}{c}3120.2\\2372.6\end{array}\right)$&$\left(\begin{array}{c}2936.3\\2188.7\end{array}\right)$&$\left(\begin{array}{c}3160.6\\2413.0\end{array}\right)$\\
		$[0,\frac12](2^{+})$ &$\left(\begin{array}{c}212.9\end{array}\right)$&$\left(\begin{array}{c}212.9\end{array}\right)$&$\left(\begin{array}{c}2982.1\end{array}\right)$&$\left(\begin{array}{c}2798.2\end{array}\right)$&$\left(\begin{array}{c}3022.4\end{array}\right)$\\
		$[0,\frac12](1^{+})$ &$\left(\begin{array}{ccc}-272.4&185.6&218.7\\185.6&-135.7&-205.9\\218.7&-205.9&-8.4\end{array}\right)$&$\left(\begin{array}{c}160.2\\-14.5\\-562.3\end{array}\right)$&$\left(\begin{array}{c}2929.4\\2754.7\\2206.9\end{array}\right)$&$\left(\begin{array}{c}2745.5\\2570.8\\2022.9\end{array}\right)$&$\left(\begin{array}{c}2969.7\\2795.0\\2247.2\end{array}\right)$\\
		$[0,\frac12](0^{+})$ &$\left(\begin{array}{cc}-515.1&356.6\\356.6&-178.4\end{array}\right)$&$\left(\begin{array}{c}47.6\\-741.1\end{array}\right)$&$\left(\begin{array}{c}2816.8\\2028.1\end{array}\right)$&$\left(\begin{array}{c}2632.9\\1844.1\end{array}\right)$&$\left(\begin{array}{c}2857.1\\2068.4\end{array}\right)$
\\
\hline\hline

\multicolumn{6}{c}{$[bn(\bar{n}\bar{n})^{I_{\bar{n}\bar{n}}}]^{I}$ states} \\\hline
$[I_{\bar{n}\bar{n}},I](J^{P})$ & $\langle H_{CM} \rangle$ &Eigenvalue  &Mass&Lower limits&Upper limits\\\hline
$[1,\frac32/\frac12](2^{+})$ &$\left(\begin{array}{c}137.3\end{array}\right)$&$\left(\begin{array}{c}137.3\end{array}\right)$&$\left(\begin{array}{c}6246.7\end{array}\right)$&$\left(\begin{array}{c}6065.2\end{array}\right)$&$\left(\begin{array}{c}6277.1\end{array}\right)$\\
$[1,\frac32/\frac12](1^{+})$ &$\left(\begin{array}{ccc}-32.8&221.6&104.5\\221.6&71.5&-180.5\\104.5&-180.5&38.4\end{array}\right)$&$\left(\begin{array}{c}282.1\\112.2\\-317.2\end{array}\right)$&$\left(\begin{array}{c}6391.5\\6221.6\\5792.1\end{array}\right)$&$\left(\begin{array}{c}6210.0\\6040.0\\5610.6\end{array}\right)$&$\left(\begin{array}{c}6421.9\\6252.0\\5822.6\end{array}\right)$\\
$[1,\frac32/\frac12](0^{+})$ &$\left(\begin{array}{cc}-117.9&312.6\\312.6&78.4\end{array}\right)$&$\left(\begin{array}{c}307.9\\-347.3\end{array}\right)$&$\left(\begin{array}{c}6417.2\\5762.0\end{array}\right)$&$\left(\begin{array}{c}6235.7\\5580.5\end{array}\right)$&$\left(\begin{array}{c}6447.7\\5792.5\end{array}\right)$\\
$[0,\frac12](2^{+})$ &$\left(\begin{array}{c}186.5\end{array}\right)$&$\left(\begin{array}{c}186.5\end{array}\right)$&$\left(\begin{array}{c}6295.9\end{array}\right)$&$\left(\begin{array}{c}6114.4\end{array}\right)$&$\left(\begin{array}{c}6326.3\end{array}\right)$\\
$[0,\frac12](1^{+})$ &$\left(\begin{array}{ccc}-238.8&221.6&261.2\\221.6&-142.9&-180.5\\261.2&-180.5&-19.2\end{array}\right)$&$\left(\begin{array}{c}159.1\\32.6\\-592.6\end{array}\right)$&$\left(\begin{array}{c}6268.4\\6142.0\\5516.7\end{array}\right)$&$\left(\begin{array}{c}6086.9\\5960.5\\5335.2\end{array}\right)$&$\left(\begin{array}{c}6298.9\\6172.4\\5547.2\end{array}\right)$\\
$[0,\frac12](0^{+})$ &$\left(\begin{array}{cc}-451.5&312.6\\312.6&-156.8\end{array}\right)$&$\left(\begin{array}{c}41.4\\-649.7\end{array}\right)$&$\left(\begin{array}{c}6150.8\\5459.7\end{array}\right)$&$\left(\begin{array}{c}5969.2\\5278.2\end{array}\right)$&$\left(\begin{array}{c}6181.2\\5490.1\end{array}\right)$\\
	\hline
	\end{tabular}
\end{table}

\begin{table}[htbp]\scriptsize
	\caption{Rearrangement decays for the $cn\bar{n}\bar{n}$ and $bn\bar{n}\bar{n}$ cases. The two numbers in the parentheses for a decay channel mean dimensionless $100|\mathcal{M}|^2/{\mathcal C}^2$ and dimensional partial width, respectively. The masses and widths are presented in units of MeV.}\label{decay-Qnnn}
	\resizebox{\textwidth}{!}{
		\begin{tabular}{c|c|ccccccc|c}\hline\hline
$[I_{\bar{n}\bar{n}},I](J^{P})$&Mass&\multicolumn{7}{c}{Channels}&$\Gamma$\\ 
			\hline 

	\multicolumn{9}{c}{$[cn(\bar{n}\bar{n})^1]^{3/2}$ } \\\hline
&&$D^*\rho$&&&&&&\\
$[1,\frac32](2^{+})$&$\left[\begin{array}{c}2925.7\end{array}\right]$&$\left[\begin{array}{c}(33.3, 233.1)\end{array}\right]$&&&&&&&$\left[\begin{array}{c}233.1\end{array}\right]$\\ 
&&$D^*\rho $&$D^*\pi$&$D\rho$&&&&\\
$[1,\frac32](1^{+})$&$\left[\begin{array}{c}3047.0\\2842.9\\2464.7 \end{array}\right]$&$\left[\begin{array}{c}(44.7, 400.2)\\(3.8, 17.9)\\(1.5, -) \end{array}\right]$&$\left[\begin{array}{c}(0.2, 2.4)\\(0.3, 4.1)\\(41.2, 392.0) \end{array}\right]$&$\left[\begin{array}{c}(9.4, 104.9)\\(29.8, 261.7)\\(2.5, -) \end{array}\right]$&&&&&$\left[\begin{array}{c}507.5\\283.7\\392.0\end{array}\right]$\\
&&$D^*\rho$&$D\pi$&&&&&\\
$[1,\frac32](0^{+})$&$\left[\begin{array}{c}3120.2\\2372.6\end{array}\right]$&$\left[\begin{array}{c}(54.9, 535.3)\\(3.5, -)\end{array}\right]$&$\left[\begin{array}{c}(0.1, 1.6)\\(41.6, 470.2)\end{array}\right]$&&&&&&$\left[\begin{array}{c}536.9\\470.2\end{array}\right]$\\
\hline			
			
	\multicolumn{9}{c}{$[cn(\bar{n}\bar{n})^1]^{1/2}$ } \\\hline
			&&$D^*\rho$&$D^*\omega$&&&&&&\\ 
			$[1,\frac12](2^{+})$&$\left[\begin{array}{c}2925.7\end{array}\right]$&$\left[\begin{array}{c}(8.3, 58.3)\end{array}\right]$&$\left[\begin{array}{c}(25.0, 170.6)\end{array}\right]$&&&&&&$\left[\begin{array}{c}228.8\end{array}\right]$\\ 
			&&$D^*\rho$&$D^*\pi$&$ D\rho$ & $ D^*\omega$&$D^*\eta$& $D^*\eta^{\prime}$&$D\omega$&\\ 
			$[1,\frac12](1^{+})$&$\left[\begin{array}{c}3047.0\\2842.9\\2464.7\end{array}\right]$&$\left[\begin{array}{c}(11.2, 100.1)\\(0.9, 4.5)\\(0.4, -)\end{array}\right]$&$\left[\begin{array}{c}(0.0, 0.6)\\(0.1, 1.0)\\(10.3, 98.0)\end{array}\right]$&$\left[\begin{array}{c}(2.3, 26.2)\\(7.4, 65.4)\\(0.6,-)\end{array}\right]$&$\left[\begin{array}{c}(33.5, 296.4)\\(2.8, 12.6)\\(1.1,-)\end{array}\right]$&$\left[\begin{array}{c}(0.1, 0.5)\\(0.1, 0.7)\\(16.3,-)\end{array}\right]$&$\left[\begin{array}{c}(0.1, 0.4)\\(0.3,-)\\(32.0,-)\end{array}\right]$&$\left[\begin{array}{c}(7.0, 78.1)\\(22.3, 193.0)\\(1.9,-)\end{array}\right]$&$\left[\begin{array}{c}502.3\\277.2\\98.0 \end{array}\right]$\\ 
			&&$D^*\rho$&$D\pi$&$D^*\omega$&$D\eta$&$D\eta^{\prime}$&&&\\ 
			$[1,\frac12](0^{+})$&$\left[\begin{array}{c}3120.2\\2372.6\end{array}\right]$&$\left[\begin{array}{c}(13.7, 133.8)\\(0.9, -) \end{array}\right]$&$\left[\begin{array}{c}(0.0, 0.4)\\(10.4, 117.5) \end{array}\right]$&$\left[\begin{array}{c}(41.2, 397.7)\\(2.6, -) \end{array}\right]$&$\left[\begin{array}{c}(0.0, 0.3)\\(16.5, -) \end{array}\right]$&$\left[\begin{array}{c}(0.1, 0.4)\\(32.3, -) \end{array}\right]$&&&$\left[\begin{array}{c}532.6\\117.5 \end{array}\right]$\\ \hline
	\multicolumn{9}{c}{$[cn(\bar{n}\bar{n})^0]^{1/2}$} \\\hline
			&&$D^*\rho$&$D^*\omega$&&&&&&\\ 
			$[0,\frac12](2^{+})$&$\left[\begin{array}{c}2982.1\end{array}\right]$&$\left[\begin{array}{c}(50.0, 401.5)\end{array}\right]$&$\left[\begin{array}{c}(16.7, 131.6)\end{array}\right]$&&&&&&$\left[\begin{array}{c}533.0\end{array}\right]$\\ 
			&&$D^*\rho$&$D^*\pi$&$ D\rho$ & $ D^*\omega$&$D^*\eta$& $D^*\eta^{\prime}$&$D\omega$&\\ 
			$[0,\frac12](1^{+})$&$\left[\begin{array}{c}2929.4\\2754.7\\2206.9\end{array}\right]$&$\left[\begin{array}{c}(34.6, 244.7)\\(2.7, -)\\(0.2, -)\end{array}\right]$&$\left[\begin{array}{c}(0.5, 6.6)\\(0.8, 10.3)\\(42.4, 173.6)\end{array}\right]$&$\left[\begin{array}{c}(9.9, 99.5)\\(33.6, 232.0)\\(0.2,-)\end{array}\right]$&$\left[\begin{array}{c}(11.5, 79.6)\\(0.9,-)\\(0.1,-)\end{array}\right]$&$\left[\begin{array}{c}(0.1, 0.5)\\(0.1, 0.6)\\(7.5,-)\end{array}\right]$&$\left[\begin{array}{c}(0.1,-)\\(0.1,-)\\(6.7,-)\end{array}\right]$&$\left[\begin{array}{c}(3.3, 32.8)\\(11.2, 74.9)\\(0.1,-)\end{array}\right]$&$\left[\begin{array}{c}463.6 \\317.8 \\173.6\end{array}\right]$\\ 
			&&$D^*\rho$&$D\pi$&$D^*\omega$&$D\eta$&$D\eta^{\prime}$&&&\\ 
			$[0,\frac12](0^{+})$&$\left[\begin{array}{c}2816.8\\2028.1\end{array}\right]$&$\left[\begin{array}{c}(30.8, 110.0)\\(0.4,-) \end{array}\right]$&$\left[\begin{array}{c}(1.4, 19.6)\\(42.4, 119.7) \end{array}\right]$&$\left[\begin{array}{c}(10.3, 32.4)\\(0.1,-) \end{array}\right]$&$\left[\begin{array}{c}(0.2, 1.4)\\(7.5,-) \end{array}\right]$&$\left[\begin{array}{c}(0.2, -)\\(6.7,-) \end{array}\right]$&&&$\left[\begin{array}{c}163.5\\119.7\end{array}\right]$\\ 
			\hline
			
	\multicolumn{9}{c}{$[bn(\bar{n}\bar{n})^1]^{3/2}$} \\\hline
	&&$\bar{B}^*\rho$&&&&&&&\\
	$[1,\frac32](2^{+})$&$\left[\begin{array}{c}6246.7\end{array}\right]$&$\left[\begin{array}{c}(33.3, 57.8)\end{array}\right]$&&&&&&&$\left[\begin{array}{c}57.8\end{array}\right]$\\
	&&$\bar{B}^*\rho $&$\bar{B}^*\pi$&$\bar{B}\rho$&&&&&\\
	$[1,\frac32](1^{+})$&$\left[\begin{array}{c}6391.5\\6221.6\\5792.1 \end{array}\right]$&$\left[\begin{array}{c}(40.0, 96.2)\\(8.0, 12.7)\\(2.0, -) \end{array}\right]$&$\left[\begin{array}{c}(0.1, 0.4)\\(0.0, 0.1)\\(41.5, 77.8) \end{array}\right]$&$\left[\begin{array}{c}(14.8, 38.6)\\(25.4, 47.5)\\(1.5, -) \end{array}\right]$&&&&&$\left[\begin{array}{c}135.2\\60.3\\77.8\end{array}\right]$\\
	&&$\bar{B}^*\rho$&$\bar{B}\pi$&&&&&\\
	$[1,\frac32](0^{+})$&$\left[\begin{array}{c}6417.2\\5762.0\end{array}\right]$&$\left[\begin{array}{c}(54.9, 137.4)\\(3.5, -)\end{array}\right]$&$\left[\begin{array}{c}(0.1, 0.4)\\(41.6, 81.3)\end{array}\right]$&&&&&&$\left[\begin{array}{c}137.8\\81.3\end{array}\right]$\\
	\hline
	
	\multicolumn{9}{c}{$[bn(\bar{n}\bar{n})^1]^{1/2}$} \\\hline
			&&$\bar{B}^*\rho$&$\bar{B}^*\omega$&&&&&&\\ 
			$[1,\frac12](2^{+})$&$\left[\begin{array}{c}6246.7\end{array}\right]$&$\left[\begin{array}{c}(8.3, 14.5)\end{array}\right]$&$\left[\begin{array}{c}(25.0, 42.3)\end{array}\right]$&&&&&&$\left[\begin{array}{c}56.8\end{array}\right]$\\ 
			&&$\bar{B}^*\rho$&$\bar{B}^*\pi$&$\bar{B}\rho$ & $ \bar{B}^*\omega$&$\bar{B}^*\eta$& $\bar{B}^*\eta^{\prime}$&$\bar{B}\omega$&\\ 
			$[1,\frac12](1^{+})$&$\left[\begin{array}{c}6391.5\\6221.6\\5792.1\end{array}\right]$&$\left[\begin{array}{c}(10.0, 24.0)\\(2.0, 3.2)\\(0.5, -)\end{array}\right]$&$\left[\begin{array}{c}(0.0, 0.1)\\(0.0, 0.0)\\(10.4, 19.5)\end{array}\right]$&$\left[\begin{array}{c}(3.7, 9.7)\\(6.3, 11.9)\\(0.4,-)\end{array}\right]$&$\left[\begin{array}{c}(30.0, 71.3)\\(6.0, 9.2)\\(1.5,-)\end{array}\right]$&$\left[\begin{array}{c}(0.0, 0.1)\\(0.0, 0.0)\\(16.5,-)\end{array}\right]$&$\left[\begin{array}{c}(0.0, 0.0)\\(0.0,-)\\(14.7,-)\end{array}\right]$&$\left[\begin{array}{c}(11.1, 28.7)\\(19.0, 34.9)\\(1.1,-)\end{array}\right]$&$\left[\begin{array}{c}133.9\\59.2 \\19.5\end{array}\right]$\\ 
			&&$\bar{B}^*\rho$&$\bar{B}\pi$&$\bar{B}^*\omega$&$\bar{B}\eta$&$\bar{B}\eta^{\prime}$&&&\\ 
			$[1,\frac12](0^{+})$&$\left[\begin{array}{c}6417.2\\5762.0\end{array}\right]$&$\left[\begin{array}{c}(13.7, 34.4)\\(0.9, -) \end{array}\right]$&$\left[\begin{array}{c}(0.0, 0.1)\\(10.4, 20.3) \end{array}\right]$&$\left[\begin{array}{c}(41.2, 102.0)\\(2.6, -) \end{array}\right]$&$\left[\begin{array}{c}(0.0, 0.1)\\(16.5, -) \end{array}\right]$&$\left[\begin{array}{c}(0.0, 0.0)\\(14.7, -) \end{array}\right]$&&&$\left[\begin{array}{c}136.6\\20.3 \end{array}\right]$\\ \hline
			
	\multicolumn{9}{c}{$[bn(\bar{n}\bar{n})^0]^{1/2}$ } \\\hline 
			&&$\bar{B}^*\rho$&$\bar{B}^*\omega$&&&&&&\\ 
			$[0,\frac12](2^{+})$&$\left[\begin{array}{c}6295.9\end{array}\right]$&$\left[\begin{array}{c}(50.0, 99.7)\end{array}\right]$&$\left[\begin{array}{c}(16.7, 32.7)\end{array}\right]$&&&&&&$\left[\begin{array}{c}132.4\end{array}\right]$\\ 
			&&$\bar{B}^*\rho$&$\bar{B}^*\pi$&$ \bar{B}\rho$ & $ \bar{B}^*\omega$&$\bar{B}^*\eta$& $\bar{B}^*\eta^{\prime}$&$\bar{B}\omega$&\\ 
			$[0,\frac12](1^{+})$&$\left[\begin{array}{c}6268.4\\6142.0\\5516.7\end{array}\right]$&$\left[\begin{array}{c}(24.2, 44.9)\\(13.0, 12.2)\\(0.3, -)\end{array}\right]$&$\left[\begin{array}{c}(0.1, 0.2)\\(1.3, 3.8)\\(42.4, 26.8)\end{array}\right]$&$\left[\begin{array}{c}(25.0, 52.6)\\(18.6, 25.4)\\(0.2,-)\end{array}\right]$&$\left[\begin{array}{c}(8.1, 14.7)\\(4.3, 3.7)\\(0.1,-)\end{array}\right]$&$\left[\begin{array}{c}(0.0, 0.0)\\(0.2, 0.3)\\(7.5,-)\end{array}\right]$&$\left[\begin{array}{c}(0.0,-)\\(0.2,-)\\(6.7,-)\end{array}\right]$&$\left[\begin{array}{c}(8.3, 17.3)\\(6.2, 8.1)\\(0.1,-)\end{array}\right]$&$\left[\begin{array}{c}129.7 \\53.4 \\26.8 \end{array}\right]$\\ 
			&&$\bar{B}^*\rho$&$\bar{B}\pi$&$\bar{B}^*\omega$&$\bar{B}\eta$&$\bar{B}\eta^{\prime}$&&&\\ 
			$[0,\frac12](0^{+})$&$\left[\begin{array}{c}6150.8\\5459.7\end{array}\right]$&$\left[\begin{array}{c}(30.8, 31.7)\\(0.4, -) \end{array}\right]$&$\left[\begin{array}{c}(1.4, 4.2)\\(42.4, 23.8) \end{array}\right]$&$\left[\begin{array}{c}(10.3, 9.8)\\(0.1, -) \end{array}\right]$&$\left[\begin{array}{c}(0.2, 0.3)\\(7.5, -) \end{array}\right]$&$\left[\begin{array}{c}(0.2, -)\\(6.7, -) \end{array}\right]$&&&$\left[\begin{array}{c}46.0 \\23.8\end{array}\right]$\\ 
			\hline\hline 
	\end{tabular}}
\end{table}

The predicted masses of the $cn\bar{n}\bar{n}$ states are listed in Table \ref{spectrum-Qnnn} and their relative positions are illustrated in Fig.\ref{relative}(a). There exist six tetraquark states for each of $[cn(\bar{n}\bar{n})^1]^\frac32$, $[cn(\bar{n}\bar{n})^1]^\frac12$, and $[cn(\bar{n}\bar{n})^0]^\frac12$. Their masses range from 2028 MeV to 3120 MeV. Note that the $[I_{\bar{n}\bar{n}},I]=[1,\frac32]$ and $[I_{\bar{n}\bar{n}},I]=[1,\frac12]$ states with the same angular momentum are degenerate in the adopted model. The $[cn(\bar{n}\bar{n})^1]^\frac32$ states do not mix with the conventional mesons. The $[cn\bar{n}\bar{n}]^\frac12$ states are generally heavier than the observed $c\bar{n}$ mesons \cite{ParticleDataGroup:2024cfk}, but the mixing between these two configurations is allowed, in principle. Here we do not consider the mixing. We present the decay information of all the $cn\bar{n}\bar{n}$ tetraquarks in Table \ref{decay-Qnnn}. It is clear that the isospin-degenerate states have different decay properties.

For the $[I_{\bar{n}\bar{n}},I]=[1,\frac32]$ case,  one finds that the lowest state has a mass around 2.4 GeV with $J^P=0^+$. 
As shown in Table \ref{spectrum-Qnnn} and Fig. \ref{relative}(a),
this state has only one rearrangement decay channel $D\pi$. The mass of another higher $0^+$ state is located around 3.1 GeV, which mainly decays into $D^*\rho$. The highest $1^+$ state $\tilde{T}(3047)$ has two dominant rearrangement decay channels $D^*\rho$ and $D\rho$, while the channel $D^*\pi$ has a suppressed contribution. The partial width ratio between $D^*\rho$ and $D\rho$ is about $\Gamma(D^*\rho):\Gamma(D\rho)\simeq3.8$.	
The decay of the second highest $1^+$ state $\tilde{T}(2843)$ mainly gets contribution from $D\rho$, while the decay channels $D^*\rho$ and $D^*\pi$  are also allowed.  The partial width ratio is found to be $\Gamma(D^*\rho):\Gamma(D\rho)\simeq0.07$.
The remaining $1^+$ ($2^+$) state has only one rearrangement decay channel $D^*\pi$ ($D^*\rho$).

For the $[I_{\bar{n}\bar{n}},I]=[1,\frac12]$ case, we just focus on the rearrangement decays listed in Table \ref{decay-Qnnn}. The lowest $J^P=0^+$ state has only one decay channel $D\pi$. Compared with its isospin-degenerate state, the width is much smaller. The other $0^+$ state mainly decays into $D^*\rho$ and $D^*\omega$, while the decays into $D\pi$, $D\eta$, and $D\eta^{\prime}$ are suppressed. The ratio between the dominant partial widths of this state is predicted to be $\Gamma(D^*\rho):\Gamma(D^*\omega)\sim 0.34$. Compared with its isospin-degenerate state, the partial width of $D^*\rho$ is smaller but the number of channels is larger, which gives finally comparable total widths.
As for the highest $1^+$ state, there are seven decay channels. The main contributions to its width are from the $D^*\rho$, $D \rho$, $D^*\omega$, and $D\omega$ channels. The ratios between their partial widths are found to be $\Gamma(D^*\rho):\Gamma(D\rho):\Gamma(D^*\omega):\Gamma(D\omega)=3.8:1.0:11.3:3.0$. 
The second highest $1^+$ state mainly decays into $D\rho$, $D^*\omega$, and $D\omega$ with the partial width ratio $\Gamma(D\rho):\Gamma(D^*\omega):\Gamma(D\omega)=5.2:1.0:15.3$. 
The lowest $1^+$ state is the narrowest $[cn(\bar{n}\bar{n})^1]^\frac12$, which has only one rearrangement decay mode $D^*\pi$. For the $2^{+}$ state, the channels $D^*\rho$ and $D^*\omega$ provide important contributions to the decay width. Their partial width ratio is estimated to be $\Gamma(D^*\rho):\Gamma(D^*\omega)\sim0.34$.

From Table \ref{spectrum-Qnnn}, we can find that the six $[I_{\bar{n}\bar{n}},I]=[0,\frac12]$ states are located at $2028\sim 2982$ MeV with widths ranging from approximately 120 MeV to 533 MeV.
The decay modes for a $[I_{\bar{n}\bar{n}},I]=[0,\frac12]$ state are the same as those for its corresponding $[I_{\bar{n}\bar{n}},I]=[1,\frac12]$ state. However, the partial width ratios of the dominant channels differ. The lower $J^P=0^+$ $[cn(\bar{n}\bar{n})^0]^\frac12$ state $\tilde{T}(2028)$ has the same rearrangement decay channel $D\pi$ as the lower $J^P=0^+$ $[cn(\bar{n}\bar{n})^1]^\frac12$. Their coupling amplitudes with $D\pi$ are different ($[cn(\bar{n}\bar{n})^1]^\frac12$ is smaller). Since the lower $J^P=0^+$ $[cn(\bar{n}\bar{n})^1]^\frac12$ state possesses a larger decay phase space than the  $\tilde{T}(2028)$, their decay widths are finally comparable. The other $J^P=0^+$ $[cn(\bar{n}\bar{n})^0]^\frac12$ state $\tilde{T}(2817)$ mainly decays into $D^*\rho$, $D\pi$, and $D^*\omega$ with the partial width ratio $\Gamma(D^*\rho):\Gamma(D\pi):\Gamma(D^*\omega)=5.6:1.0:1.6$. The first channel is its main decay mode, while the last one is the main decay mode for the higher $0^+$ $[cn(\bar{n}\bar{n})^1]^\frac12$ state.
The highest $J^P=1^+$ $[cn(\bar{n}\bar{n})^0]^\frac12$ state $\tilde{T}(2929)$ and the highest $1^+$ $[cn(\bar{n}\bar{n})^1]^\frac12$ state have the same main rearrangement decay channels, but their partial width ratios are significantly different. For the $\tilde{T}(2929)$, the ratios are about  $\Gamma(D^*\rho):\Gamma(D\rho):\Gamma(D^*\omega):\Gamma(D\omega)=7.5:3.0:2.4:1.0$.
 The second highest $J^P=1^+$ $[cn(\bar{n}\bar{n})^0]^\frac12$ state $\tilde{T}(2755)$ mainly decays into $D^*\pi$, $D\rho$, and $D\omega$ channels. The related ratios are predicted to be $\Gamma(D^*\pi):\Gamma(D\rho):\Gamma(D\omega)=1.0:22.5:7.3$, which differ from the second highest $1^+$ $[cn(\bar{n}\bar{n})^1]^\frac12$. The decay feature for the lightest $J^P=1^+$ $[cn(\bar{n}\bar{n})^0]^\frac12$ state $\tilde{T}(2207)$ is similar to the lightest $1^+$ $[cn(\bar{n}\bar{n})^1]^\frac12$. The former state is below the latter one, but its strong coupling with their unique rearrangement decay channel $D^*\pi$ leads to a larger width.
 The decay of $J^P=2^+$ $[cn(\bar{n}\bar{n})^0]^\frac12$ state $\tilde{T}(2982)$ gets contribution mainly from $D^*\rho$ and $D^*\omega$. Their partial width ratio is predicted to be $\Gamma(D^*\rho):\Gamma(D^*\omega)=3.1$, which is larger than the ratio for the $2^+$ $[cn(\bar{n}\bar{n})^1]^\frac12$ state.\\

One may similarly analyze the $bn\bar{n}\bar{n}$ states. We also list the obtained values for the mass spectrum and the rearrangement decay widths for this case in Tables \ref{spectrum-Qnnn} and \ref{decay-Qnnn}, respectively. The relative positions for the involved states are shown in Fig. \ref{relative}(g). The mass ordering and the relative splitting of the $bn\bar{n}\bar{n}$ system exhibit strong similarities to those of the $cn\bar{n}\bar{n}$ system. Using the same decay parameter ${\cal C}$ as the charmed case, one reaches a conclusion that a bottom tetraquark is narrower than its corresponding charmed state. If this parameter is twice as large in the bottom case as in the charmed case, the decay widths are then comparable.

From Table \ref{spectrum-Qnnn} or Fig.\ref{relative} (g), the estimated masses for the $[bn(\bar{n}\bar{n})^1]^\frac32$ states are around 5762$\sim$6392 MeV, and  they are all unstable. As shown in Table \ref{decay-Qnnn}, the lower $J^P=0^+$ $[bn(\bar{n}\bar{n})^1]^\frac32$ state and the higher one dominantly decay into $\bar{B}\pi$ and $\bar{B}^*\rho$, respectively. For the $J^P=1^+$ $[bn(\bar{n}\bar{n})^1]^\frac32$ tetraquarks, the highest and the second highest state share the main decay modes $\bar{B}^*\rho$ and $\bar{B}\rho$, while the lightest state mainly decays into $\bar{B}^*\pi$. The observation of the tensor $[bn(\bar{n}\bar{n})^1]^\frac32$ state in the $\bar{B}^*\rho$ channel is possible.
 
For two degenerate $[bn(\bar{n}\bar{n})^1]^\frac12$ and $[bn(\bar{n}\bar{n})^1]^\frac32$ states with the same $J$, the $I=1/2$ state has more rearrangement decay channels. While its coupling strengths with the common channels are weaker than the $I=3/2$ state, the larger number of available channels compensates for this. As a result, its total width is comparable to that of the $I=3/2$ state. The exception cases are the lowest $J^P=0^+$ and $1^+$ states. For the $I=1/2$ states, the number of allowed decay channels is kinematically reduced compared to their $I=3/2$ counterparts. Their widths are therefore much narrower. 

The predicted masses of the $[bn(\bar{n}\bar{n})^0]^\frac12$ states are around 5460$\sim$6296 MeV. From Table \ref{decay-Qnnn}, the lower $J^P=0^+$ $[bn(\bar{n}\bar{n})^0]^\frac12$ and $[bn(\bar{n}\bar{n})^1]^\frac12$ states can be searched for in the $\bar{B}\pi$ channel. In the higher $J^P=0^+$ case, the channel $\bar{B}^*\omega$ is more strongly coupled with the $[bn(\bar{n}\bar{n})^1]^\frac12$ state than the $[bn(\bar{n}\bar{n})^0]^\frac12$ state, while the channel $\bar{B}^*\rho$ exhibits the opposite trend. Because of its lower mass, the higher $J^P=0^+$ $[bn(\bar{n}\bar{n})^0]^\frac12$ state has a narrower width than the higher $0^+$ $[bn(\bar{n}\bar{n})^1]^\frac12$.
Both the lowest $J^P=1^+$ $[bn(\bar{n}\bar{n})^0]^\frac12$ and the lowest $1^+$ $[bn(\bar{n}\bar{n})^1]^\frac12$ states have only one two-body decay channel $\bar{B}^*\pi$. The stronger coupling makes the former state a slightly broader one, although its mass is smaller. To search for other $J^P=1^+$ states (for both $[bn(\bar{n}\bar{n})^0]^\frac12$ and $[bn(\bar{n}\bar{n})^1]^\frac12$ cases), the $\bar{B}^*\rho$, $\bar{B}\rho$, $\bar{B}^*\omega$, and $\bar{B}\omega$ channels can be employed. 
 For the $J^P=2^+$ states, the rearrangement decay in the $[bn(\bar{n}\bar{n})^0]^\frac12$ case gets major contribution from the $\bar{B}^*\rho$ channel, whereas the dominant contribution for the decay in the $[bn(\bar{n}\bar{n})^1]^\frac12$ case comes from the channel $\bar{B}^*\omega$.

\subsection{The $cn\bar{s}\bar{n}$ and $bn\bar{s}\bar{n}$ systems}\label{sec-Tcsbar}

\begin{table}[htbp]
	\caption{Mass spectra for the $cn\bar{s}\bar{n}$ and $bn\bar{s}\bar{n}$ states in units of MeV. The thresholds of $D_s\pi$ and $\bar{B}_s\pi$ are used to establish the lower limits for the tetraquark masses in the former and latter cases, respectively.  The upper limits are obtained using the effective quark masses: $m_n=361.8$ MeV, $m_s=542.4$ MeV, $m_c=1724.1$ MeV, and $m_b=5054.4$ MeV. We adopt values from the mass splitting model (fourth column) in the discussions. The systems have degenerate states with different isospins.}\label{spectrum-Qnsn}\scriptsize
	\begin{tabular}{c|ccccccc}\hline
		\hline\multicolumn{6}{c}{$[cn\bar{s}\bar{n}]^I$ states} \\\hline
		$I(J^{P})$ & $\langle H_{CMI} \rangle$ &Eigenvalue  &Mass&Lower limits&Upper limits\\\hline
		$1/0(2^+)$ &$\left(\begin{array}{cc}184.5&-31.7\\-31.7&125.3\end{array}\right)$&$\left(\begin{array}{c}198.3\\111.6\end{array}\right)$&$\left(\begin{array}{c}3058.0\\2971.3\end{array}\right)$&$\left(\begin{array}{c}2888.6\\2801.9\end{array}\right)$&$\left(\begin{array}{c}3188.4\\3101.7\end{array}\right)$\\
		$1/0(1^{+})$ &$\left(\begin{array}{cccccc}-227.5&31.7&-51.9&140.8&165.9&-44.0\\31.7&-39.5&140.8&-20.7&-44.0&66.4\\-51.9&140.8&43.1&0&37.3&-174.8\\140.8&-20.7&0&-86.1&-174.8&14.9\\165.9&-44.0&37.3&-174.8&-0.1&0\\-44.0&66.4&-174.8&14.9&0&0.3\end{array}\right)$&$\left(\begin{array}{c}222.4\\144.6\\42.9\\-22.5\\-223.4\\-473.9\end{array}\right)$&$\left(\begin{array}{c}3082.1\\3004.4\\2902.7\\2837.3\\2636.4\\2385.9\end{array}\right)$&$\left(\begin{array}{c}2912.7\\2834.9\\2733.3\\2667.9\\2467.0\\2216.4\end{array}\right)$&$\left(\begin{array}{c}3212.5\\3134.7\\3033.0\\2967.6\\2766.7\\2516.2\end{array}\right)$\\
		$1/0(0^{+})$ &$\left(\begin{array}{cccc}-433.5&63.4&-64.7&302.8\\63.4&-121.9&302.8&-25.9\\-64.7&302.8&64.4&0\\302.8&-25.9&0&-128.8\end{array}\right)$&$\left(\begin{array}{c}289.9\\58.1\\-322.3\\-645.4\end{array}\right)$&$\left(\begin{array}{c}3149.7\\2917.9\\2537.5\\2214.3\end{array}\right)$&$\left(\begin{array}{c}2980.2\\2748.4\\2368.0\\2044.9\end{array}\right)$&$\left(\begin{array}{c}3280.0\\3048.2\\2667.8\\2344.7\end{array}\right)$
		\\\hline
		\hline\multicolumn{6}{c}{$[bn\bar{s}\bar{n}]^I$ states} \\\hline
		$I(J^{P})$ & $\langle H_{CM} \rangle$ &Eigenvalue  &Mass&Lower limits&Upper limits\\\hline
		$1/0(2^{+})$ &$\left(\begin{array}{cc}158.5&-32.0\\-32.0&106.3\end{array}\right)$&$\left(\begin{array}{c}173.6\\91.1\end{array}\right)$&$\left(\begin{array}{c}6373.6\\6291.1\end{array}\right)$&$\left(\begin{array}{c}6192.1\\6109.5\end{array}\right)$&$\left(\begin{array}{c}6494.0\\6411.5\end{array}\right)$\\
		$1/0(1^{+})$ &$\left(\begin{array}{cccccc}-194.2&32.0&-51.4&176.4&207.9&-43.6\\32.0&-34.8&176.4&-20.6&-43.6&83.2\\-51.4&176.4&46.7&0&37.7&-149.6\\176.4&-20.6&0&-93.3&-149.6&15.1\\207.9&-43.6&37.7&-149.6&-10.9&0\\-43.6&83.2&-149.6&15.1&0&21.9\end{array}\right)$&$\left(\begin{array}{c}222.0\\144.0\\69.7\\37.6\\-236.6\\-501.5\end{array}\right)$&$\left(\begin{array}{c}6422.0\\6344.0\\6269.6\\6237.6\\5963.3\\5698.5\end{array}\right)$&$\left(\begin{array}{c}6240.5\\6162.4\\6088.1\\6056.0\\5781.8\\5517.0\end{array}\right)$&$\left(\begin{array}{c}6542.4\\6464.4\\6390.1\\6358.0\\6083.8\\5818.9\end{array}\right)$\\
		$1/0(0^{+})$ &$\left(\begin{array}{cccc}-370.5&63.9&-65.2&259.2\\63.9&-105.3&259.2&-26.1\\-65.2&259.2&53.6&0\\259.2&-26.1&0&-107.2\end{array}\right)$&$\left(\begin{array}{c}247.5\\52.1\\-269.9\\-559.2\end{array}\right)$&$\left(\begin{array}{c}6447.4\\6252.1\\5930.1\\5640.8\end{array}\right)$&$\left(\begin{array}{c}6265.9\\6070.6\\5748.6\\5459.3\end{array}\right)$&$\left(\begin{array}{c}6567.9\\6372.5\\6050.5\\5761.2\end{array}\right)$\\
		\hline
	\end{tabular}
\end{table}

\begin{table}[!h]\scriptsize
	\caption{Rearrangement decays for the $cn\bar{s}\bar{n}$ and $bn\bar{s}\bar{n}$ cases. The two numbers in the parentheses for a decay channel mean dimensionless $100|\mathcal{M}|^2/{\mathcal C}^2$ and dimensional partial width, respectively. The masses and widths are presented in units of MeV.}\label{decay-Qnsn}
	\resizebox{\textwidth}{!}{\begin{tabular}{c|c|ccccccc|c}\hline\hline
			
$I(J^{P})$&Mass&\multicolumn{7}{c}{Channels}&$\Gamma$\\ \hline 
\multicolumn{9}{c}{$cn\bar{s}\bar{n}$ } \\\hline
&&$D^*_s\rho$&$D^*K^*$&&&&&\\
$1(2^{+})$&$\left[\begin{array}{c}3058.0\\2971.3\end{array}\right]$&$\left[\begin{array}{c}(95.8, 340.4)\\(4.2, 10.9)\end{array}\right]$&$\left[\begin{array}{c}(26.9, 94.8)\\(73.1, 179.3)\end{array}\right]$&&&&&&$\left[\begin{array}{c}435.2\\190.1\end{array}\right]$\\
&&$D_s^*\rho$&$D_s^*\pi$&$D_s\rho$&$D^*K^*$&$D^*K$&$DK^*$&\\
$1(1^{+})$&$\left[\begin{array}{c}3082.1\\3004.4\\2902.7\\2837.3\\2636.4\\2385.9\end{array}\right]$&$\left[\begin{array}{c}(71.1, 266.6)\\(26.8, 80.9)\\(0.0, 0.0)\\(1.3, -)\\(0.8, -)\\(0.0, -)\end{array}\right]$&$\left[\begin{array}{c}(0.0, 0.2)\\(0.3, 2.0)\\(0.7, 4.2)\\(0.0, 0.0)\\(10.7, 51.5)\\(88.2, 253.0)\end{array}\right]$&$\left[\begin{array}{c}(4.5, 22.7)\\(22.7, 103.6)\\(44.1, 166.0)\\(26.8, 80.2)\\(1.8, -)\\(0.0, -)\end{array}\right]$&$\left[\begin{array}{c}(22.5, 84.1)\\(61.8, 181.3)\\(12.6, 2.7)\\(0.1, -)\\(1.5,-)\\(1.4, -)\end{array}\right]$&$\left[\begin{array}{c}(0.8, 4.6)\\(1.3, 7.5)\\(0.0, 0.2)\\(3.2, 16.5)\\(72.7, 258.8)\\(22.1, -)\end{array}\right]$&$\left[\begin{array}{c}(9.3, 46.7)\\(1.3, 6.1)\\(12.4, 45.5)\\(70.9, 198.4)\\(4.4,-)\\(1.7, -)\end{array}\right]$&&$\left[\begin{array}{c}424.9\\381.4\\218.5\\295.1\\310.3\\253.0\end{array}\right]$\\
&&$D_s^*\rho$&$D_s\pi$&$D^*K^*$&$DK$&&&\\
$1(0^{+})$&$\left[\begin{array}{c}3149.7\\2917.9\\2537.5\\2214.3\end{array}\right]$&$\left[\begin{array}{c}(63.3, 266.6)\\(34.1, 54.9)\\(2.5, -)\\(0.1, -)\end{array}\right]$&$\left[\begin{array}{c}(0.0, 0.3)\\(0.7, 5.1)\\(17.3, 96.4)\\(81.9, 235.6)\end{array}\right]$&$\left[\begin{array}{c}(46.3, 196.1)\\(47.4, 56.9)\\(4.1, -)\\(2.1, -)\end{array}\right]$&$\left[\begin{array}{c}(0.2, 1.7)\\(3.1, 20.1)\\(67.0, 297.1)\\(29.6, -)\end{array}\right]$&&&&$\left[\begin{array}{c}464.6\\137.0\\393.5\\235.6\end{array}\right]$\\
\hline

     \multicolumn{9}{c}{$cn\bar{s}\bar{n}$ } \\\hline
			&&$D^*_s\omega$&$D^*K^*$&&&&&&\\ 
			$0(2^{+})$&$\left[\begin{array}{c}3058.0\\2971.3\end{array}\right]$&$\left[\begin{array}{c}(95.8, 333.6)\\(4.2, 10.4)\end{array}\right]$&$\left[\begin{array}{c}(26.9, 94.8)\\(73.1, 179.3)\end{array}\right]$&&&&&&$\left[\begin{array}{c}428.4\\189.7\end{array}\right]$\\ 
			&&$D_s^*\omega$&$D_s^*\eta$&$D_s^*\eta^{\prime}$&$D_s\omega$&$D^*K^*$&$D^*K$&$DK^*$&\\ 
			$0(1^{+})$&$\left[\begin{array}{c}3082.1\\3004.4\\2902.7\\2837.3\\2636.4\\2385.9\end{array}\right]$&$\left[\begin{array}{c}(71.1, 262.0)\\(26.8, 78.5)\\(0.0, 0.0)\\(1.3, -)\\(0.8, -)\\(0.0,-)\end{array}\right]$&$\left[\begin{array}{c}(0.0, 0.1)\\(0.2, 0.8)\\(0.4, 1.6)\\(0.4,0.0)\\(5.7, -)\\(46.7, -)\end{array}\right]$&$\left[\begin{array}{c}(0.0, 0.0)\\(0.2,-)\\(0.3, -)\\(0.0, -)\\(5.0, -)\\(41.5, -)\end{array}\right]$&$\left[\begin{array}{c}(4.5, 22.5)\\(22.7, 102.3)\\(44.1, 162.4)\\(26.8, 77.1)\\(1.8, -)\\(0.0, -)\end{array}\right]$&$\left[\begin{array}{c}(22.5, 84.1)\\(61.8, 181.3)\\(12.6, 2.7)\\(0.1, -)\\(1.5, -)\\(1.4, -)\end{array}\right]$&$\left[\begin{array}{c}(0.8, 4.6)\\(1.3, 7.5)\\(0.0, 0.2)\\(3.2, 16.5)\\(72.7, 258.8)\\(22.1, -)\end{array}\right]$&$\left[\begin{array}{c}(9.3, 46.7)\\(1.3, 6.1)\\(12.4, 45.5)\\(70.9, 198.4)\\(4.4, -)\\(1.7, -)\end{array}\right]$&$\left[\begin{array}{c}420.0\\376.5 \\212.3 \\292.0 \\258.8 \\-\end{array}\right]$\\ 
			&&$D_s^*\omega$&$D_s\eta$&$D_s\eta^{\prime}$&$D^*K^*$&$DK$&&&\\ 
			$0(0^{+})$&$\left[\begin{array}{c}3149.7\\2917.9\\2537.5\\2214.3\end{array}\right]$&$\left[\begin{array}{c}(63.3, 263.3)\\(34.1, 47.8)\\(2.5, -)\\(0.1, -)\end{array}\right]$&$\left[\begin{array}{c}(0.0, 0.1)\\(0.4, 2.2)\\(9.1, 14.1)\\(43.3, -)\end{array}\right]$&$\left[\begin{array}{c}(0.0, 0.1)\\(0.4, 0.1)\\(8.1, -)\\(38.6, -)\end{array}\right]$&$\left[\begin{array}{c}(46.3, 196.1)\\(47.4, 56.9)\\(4.1, -)\\(2.1, -)\end{array}\right]$&$\left[\begin{array}{c}(0.2, 1.7)\\(3.1, 20.1)\\(67.0, 297.1)\\(29.6, -)\end{array}\right]$&&&$\left[\begin{array}{c}461.3\\127.1\\311.2\\-\end{array}\right]$\\ 
			\hline
			
	\multicolumn{9}{c}{$bn\bar{s}\bar{n}$} \\\hline
	&&$\bar{B}^*_s\rho$&$\bar{B}^*K^*$&&&&&\\
	$1(2^{+})$&$\left[\begin{array}{c}6373.6\\6291.1\end{array}\right]$&$\left[\begin{array}{c}(97.1, 91.1)\\(2.9, 2.1)\end{array}\right]$&$\left[\begin{array}{c}(24.0, 21.8)\\(76.0, 47.7)\end{array}\right]$&&&&&&$\left[\begin{array}{c}112.9\\49.7\end{array}\right]$\\
	&&$\bar{B}_s^*\rho$&$\bar{B}_s^*\pi$&$\bar{B}_s\rho$&$\bar{B}^*K^*$&$\bar{B}^*K$&$\bar{B}K^*$&\\
	$1(1^{+})$&$\left[\begin{array}{c}6422.0\\6344.0\\6269.6\\6237.6\\5963.3\\5698.5\end{array}\right]$&$\left[\begin{array}{c}(55.7, 58.5)\\(32.2, 27.7)\\(0.5, 0.3)\\(10.3, 5.0)\\(1.2, -)\\(0.0,-)\end{array}\right]$&$\left[\begin{array}{c}(0.0, 0.0)\\(0.0, 0.1)\\(0.2, 0.3)\\(0.4, 0.5)\\(12.6, 13.1)\\(86.8, 47.2)\end{array}\right]$&$\left[\begin{array}{c}(12.1, 14.1)\\(57.1, 57.0)\\(13.1, 10.5)\\(16.6, 11.5)\\(1.0,-)\\(0.0, -)\end{array}\right]$&$\left[\begin{array}{c}(27.8, 28.8)\\(23.8, 19.5)\\(37.0, 19.6)\\(7.3, 2.4)\\(2.4, -)\\(1.6, -)\end{array}\right]$&$\left[\begin{array}{c}(0.4, 0.6)\\(0.2, 0.3)\\(0.2, 0.3)\\(3.2, 4.2)\\(72.1, 56.7)\\(23.9, -)\end{array}\right]$&$\left[\begin{array}{c}(13.6, 15.6)\\(6.4, 6.2)\\(24.5, 17.9)\\(52.4, 31.4)\\(2.0, -)\\(1.0, -)\end{array}\right]$&&$\left[\begin{array}{c}117.6\\110.8\\48.9\\55.0\\69.8\\47.2\end{array}\right]$\\
	&&$\bar{B}_s^*\rho$&$\bar{B}_s\pi$&$\bar{B}^*K^*$&$\bar{B}K$&&&\\
	$1(0^{+})$&$\left[\begin{array}{c}6447.4\\6252.1\\5930.1\\5640.8\end{array}\right]$&$\left[\begin{array}{c}(65.0, 71.7)\\(32.7, 18.0)\\(2.3,-)\\(0.1, -)\end{array}\right]$&$\left[\begin{array}{c}(0.0, 0.1)\\(0.6, 1.0)\\(14.3, 15.5)\\(85.0, 45.2)\end{array}\right]$&$\left[\begin{array}{c}(44.7, 48.8)\\(48.7, 20.9)\\(4.2, -)\\(2.4, -)\end{array}\right]$&$\left[\begin{array}{c}(0.3, 0.5)\\(3.4, 4.9)\\(70.3, 58.4)\\(26.0, -)\end{array}\right]$&&&&$\left[\begin{array}{c}121.0\\44.7\\73.9\\45.2\end{array}\right]$\\
	\hline
			\multicolumn{9}{c}{$bn\bar{s}\bar{n}$ } \\\hline
			&&$\bar{B}^*_s\omega$&$\bar{B}^*K^*$&&&&&&\\ 
			$0(2^{+})$&$\left[\begin{array}{c}6373.6\\6291.1\end{array}\right]$&$\left[\begin{array}{c}(97.1, 89.4)\\(2.9, 2.0)\end{array}\right]$&$\left[\begin{array}{c}(24.0, 21.8)\\(76.0, 47.7)\end{array}\right]$&&&&&&$\left[\begin{array}{c}111.2\\49.7\end{array}\right]$\\  
			&&$\bar{B}_s^*\omega$&$\bar{B}_s^*\eta$&$\bar{B}_s^*\eta^{\prime}$&$\bar{B}_s\omega$&$\bar{B}^*K^*$&$\bar{B}^*K$&$\bar{B}K^*$&\\ 
			$0(1^{+})$&$\left[\begin{array}{c}6422.0\\6344.0\\6269.6\\6237.6\\5963.3\\5698.5\end{array}\right]$&$\left[\begin{array}{c}(55.7, 57.7)\\(32.2, 27.1)\\(0.5, 0.3)\\(10.3, 4.6)\\(1.2, -)\\(0.0, -)\end{array}\right]$&$\left[\begin{array}{c}(0.0, 0.0)\\(0.0, 0.0)\\(0.1, 0.1)\\(0.1, 0.2)\\(6.7, 0.1)\\(45.9, -)\end{array}\right]$&$\left[\begin{array}{c}(0.0, 0.0)\\(0.0, 0.0)\\(0.1, 0.1)\\(0.2, 0.2)\\(5.9, 0.1)\\(40.9, -)\end{array}\right]$&$\left[\begin{array}{c}(12.1, 14.0)\\(57.1, 56.1)\\(13.1, 10.2)\\(16.6, 11.1)\\(1.0, -)\\(0.0, -)\end{array}\right]$&$\left[\begin{array}{c}(27.8, 28.8)\\(23.8, 19.5)\\(37.0, 19.6)\\(7.3, 2.4)\\(2.4, -)\\(1.6, -)\end{array}\right]$&$\left[\begin{array}{c}(0.4, 0.6)\\(0.2, 0.3)\\(0.2, 0.3)\\(3.2, 4.2)\\(72.1, 56.7)\\(23.9, -)\end{array}\right]$&$\left[\begin{array}{c}(13.6, 15.6)\\(6.4, 6.2)\\(24.5, 17.9)\\(52.4, 31.4)\\(2.0, -)\\(1.0, -)\end{array}\right]$&$\left[\begin{array}{c}116.6\\109.3\\48.6\\54.1\\56.8\\-\end{array}\right]$\\ 
			&&$\bar{B}_s^*\omega$&$\bar{B}_s\eta$&$\bar{B}_s\eta^{\prime}$&$\bar{B}^*K^*$&$\bar{B}K$&&&\\ 
			$0(0^{+})$&$\left[\begin{array}{c}6447.4\\6252.1\\5930.1\\5640.8\end{array}\right]$&$\left[\begin{array}{c}(65.0, 70.8)\\(32.7, 16.9)\\(2.3, -)\\(0.1, -)\end{array}\right]$&$\left[\begin{array}{c}(0.0, 0.0)\\(0.3, 0.4)\\(7.6, 2.0)\\(45.0, -)\end{array}\right]$&$\left[\begin{array}{c}(0.0, 0.0)\\(0.3, 0.0)\\(6.7, -)\\(40.0, -)\end{array}\right]$&$\left[\begin{array}{c}(44.7, 48.8)\\(48.7, 20.9)\\(4.2, -)\\(2.4, -)\end{array}\right]$&$\left[\begin{array}{c}(0.3, 0.5)\\(3.4, 4.9)\\(70.3, 58.4)\\(26.0, -)\end{array}\right]$&&&$\left[\begin{array}{c}120.1\\43.1\\60.3\\ -\end{array}\right]$\\
			\hline\hline  
	\end{tabular}}
\end{table}

The $cn\bar{s}\bar{n}$ states are of special interest because they are related to the recently observed exotic $T^a_{c\bar{s}}(2900)^0$ and $T^a_{c\bar{s}}(2900)^{++}$ by the LHCb Collaboration \cite{LHCb:2022sfr, LHCb:2022lzp}. There are twelve possible  tetraquark ground  states  for $I=0$ and twelve for $I=1$. The present framework gives same-$J$ degenerate $I=0$ and $I=1$ states. For the $I=0$ case, it is possible to consider the mixing between the tetraquark and the $c\bar{s}$ configurations in future work. 
 We present the estimated $cn\bar{s}\bar{n}$ masses and their relative positions in Table \ref{spectrum-Qnsn} and Fig.\ref{relative}(b), respectively. The rearrangement decay properties of the degenerate $I=1$ and $I=0$ tetraquarks are different. They are collected in Table \ref{decay-Qnsn}.
 
For the $I=1$ case, just from the mass, the theoretical $\tilde{T}^a(2918)$ and $\tilde{T}^a(2903)$ are both consistent with the observed $T^a_{c\bar{s}0}(2900)$. 
However, the spin-parity of $T^a_{c\bar{s}0}(2900)$ determined by the LHCb is $0^+$ \cite{LHCb:2022sfr, LHCb:2022lzp}. Therefore, our results indicate that
the $\tilde{T}^a(2918)$ with $J^P=0^+$ is a good candidate for the observed $T^a_{c\bar{s}0}(2900)$. In our calculation, based on this correspondence, we determine the decay parameter ${\cal C}$.
If the $T^a_{c\bar{s}0}(2900)$ indeed corresponds to the $\tilde{T}^a(2918)$, it may be the narrowest $I=1$ $cn\bar{s}\bar{n}$ state from Table \ref{decay-Qnsn}, although it has four dominant decay modes: $D^*_s\rho$, $D_s\pi$, $D^*K^*$, and $DK$.
 The partial width ratio is found to be
\begin{eqnarray}
	\Gamma(D^*_s\rho):\Gamma(D_s\pi):\Gamma(D^*K^*):\Gamma(DK)\simeq10.8:1.0:11.2:3.9,
\end{eqnarray}
which differs from the results in the other works \cite{Lian:2023cgs,Liu:2022hbk}. This ratio can be tested by future experiments.

Our results indicate that the $\tilde{T}^a(3150)$ with $J^P=0^+$ may be the broadest  $I=1$ $cn\bar{s}\bar{n}$ state. It would predominantly decay into $D^*_s\rho$ and $D^*K^*$, while the $D_s\pi$ and $DK$ channels are suppressed by weak couplings. Our ratio between the two dominant channels is about 1.4.
The dominated rearrangement decay channels for the $\tilde{T}^a(2538)$ with $J^P=0^+$ are $D_s\pi$ and $DK$, and the corresponding partial width ratio is $\Gamma(D_s\pi):\Gamma(DK)\sim0.3$. The lightest $I(J^P)=1(0^+)$ state with mass $m=2214.3$ MeV can be searched for in its unique rearrangement decay channel $D_s\pi$.
All the six $I(J^P)=1(1^+)$ tetraquark states have the decay channel $D_s^*\pi$, which may be used to seek them. Another ideal channel for the search in experiments is the $D^*K$. Through the competition among the three effects, number of decay channels, coupling strength, and phase space, the third highest $\tilde{T}^a(2903)$ of the six $1^+$ states shows the smallest width. Both $I(J^P)=1(2^+)$ states can decay into $D^*_s\rho$ and $D^*K^*$. The former channel contributes dominantly to the width of the heavier $\tilde{T}^a(3058)$, while the latter channel contributes to the lighter $\tilde{T}^a(2971)$ dominantly.

For the $I=0$ case, the spectrum is the same as the $I=1$ case, but the primary decay channels are different. From Table \ref{decay-Qnsn}, the lowest $J^P=0^+$ $\tilde{T}^f(2214)$ should be stable since no rearrangement decay channel is allowed. This tetraquark can be searched for in its weak decay mode $K^+K^-\pi^+$. The higher $0^+$ $\tilde{T}^f(2538)$ has a main decay channel $DK$ and a suppressed channel $D_s\eta$. Their width ratio is predicted to be $\Gamma(D_s\eta):\Gamma(DK)\sim0.04$. The even higher $0^+$ $\tilde{T}^f(2918)$ dominantly decays into $D^*_s\omega$, $D^*K^*$, and $DK$. The width ratio between these channels is $\Gamma(D^*_s\omega):\Gamma(D^*K^*):\Gamma(DK)=2.4:2.8:1.0$. The highest $\tilde{T}^f(3150)$ is the broadest $I(J^P)=0(0^+)$ $cn\bar{s}\bar{n}$, which dominantly decays into $D^*_s\omega$ and $D^*K^*$. The obtained partial width ratio is about 1.3. Based on our calculation, the lowest $I(J^P)=0(1^+)$ state $\tilde{T}^f(2386)$ is a stable one. It can be searched for in the $DK\gamma$ channel. The other higher $0(1^+)$ states have rearrangement decay channels and a search for them in the $D^*K$ channel is proposed.
Both $I=0$ tensor states can decay into $D_s^*\omega$ and $D^*K^*$ channels. The corresponding partial width ratios for the higher and lower states are $\Gamma(D_s^*\omega):\Gamma(D^*K^*)\sim3.5$ and $\Gamma(D_s^*\omega):\Gamma(D^*K^*)\sim0.06$, respectively.
If the observed $T^a_{c\bar{s}0}(2900)$ is really a tetraquark, the other isovector and isoscalar $cn\bar{s}\bar{n}$ states should also exist in principle. Searching for them in various decay channels may deepen our understanding of the strong interaction.\\

The $bn\bar{s}\bar{n}$ system, the flavor partner of the $cn\bar{s}\bar{n}$ system, may be similarly analyzed. The estimated masses and relative positions of the $bn\bar{s}\bar{n}$ states are shown in Table \ref{spectrum-Qnsn} and Fig.~\ref{relative}(h), respectively. The corresponding rearrangement decay properties are also collected in Table \ref{decay-Qnsn}.
 
The $bn\bar{s}\bar{n}$ system is associated with the state $X(5568)$ observed in the $\bar{B}_s\pi$ channel by the D0 experiment \cite{D0:2016mwd}, which was not confirmed later by other experiments \cite{LHCb:2016dxl,CMS:2017hfy,CDF:2017dwr,ATLAS:2018udc}. From Table \ref{spectrum-Qnsn} and Fig. \ref{relative}(h), one finds that the mass of $X(5568)$ is about 70 MeV lower than the lowest $I(J^P)=1(0^+)$ tetraquark. If we artificially adjust the mass of the lowest state to 5568 MeV and still use the decay parameter extracted from the $T_{c\bar{s}0}^a(2900)$, the decay width would become $29.1$ MeV, which is consistent with the measured value $21.9\pm6.4^{+5.0}_{-2.5}$ MeV. The former condition implies that the uncertainty for the adopted model in predicating the $bn\bar{s}\bar{n}$ masses is about 70 MeV. The latter condition indicates that the adopted assumption ${\cal C}(bottom\,sector)\approx {\cal C}(charm\,sector)$ is valid. More studies are needed to clarify the nature of $X(5568)$ on both experimental and theoretical sides.

Of the other $J^P=0^+$ states, the lightest $\tilde{T}^f(5641)$ with $I=0$ is stable, as it has no rearrangement decay channels. It can be searched for in the $J/\psi K^-K^+$ channel. The heavier $\tilde{T}^{a}(5930)$ ($\tilde{T}^{f}(5930)$) with $I=1$ ($I=0$) mainly decays into $\bar{B}_s\pi$ and $\bar{B}K$ ($\bar{B}K$). The even heavier $\tilde{T}^{a}(6252)$ ($\tilde{T}^{f}(6252)$) with $I=1$ ($I=0$) dominantly decays into $\bar{B}_s^*\rho$, $\bar{B}^*K^*$, and $\bar{B}K$ ($\bar{B}_s^*\omega$, $\bar{B}^*K^*$, and $\bar{B}K$).
 The heaviest $\tilde{T}^a(6447)$ ($\tilde{T}^f(6447)$) with $I=1$ $(I=0)$ strongly couples to the rearrangement decay channels $\bar{B}^*_s\rho$ and $\bar{B}^*K^*$ ($\bar{B}^*_s\omega$ and $\bar{B}^*K^*$), which results in a broad total decay width.
 
In the $J^P=1^+$ case, the lightest state with $I=1$ mainly decays into $\bar{B}_s^*\pi$, whereas the lightest state with $I=0$ may be stable. This isoscalar $\tilde{T}^f(5699)$ tetraquark can be searched for in the $B^-K^+\gamma$ mode. The dominant decay modes for the second lightest state with $I=1$ are $\bar{B}_s^*\pi$ and $\bar{B}^*K$ and that for the $I=0$ is $\bar{B}^*K$. Those for the third lightest state with $I=1$ ($I=0$) are $\bar{B}^{(*)}_s\rho$, $\bar{B}^*K$, and $\bar{B}K^*$ ($\bar{B}^{(*)}_s\omega$, $\bar{B}^*K$, and  $\bar{B}K^*$). The channels $\bar{B}_s\rho$, $\bar{B}^*K^*$, and $\bar{B}K^*$ ( $\bar{B}_s\omega$, $\bar{B}^*K^*$, and $\bar{B}K^*$) provide significant contributions to the third highest state $\tilde{T}^a(6270)$ ($\tilde{T}^f(6270)$) with $I=1$ ($I=0$).
The highest $I=1$ ($I=0$) and the second highest $I=1$ ($I=0$) states share four dominant decay modes, $\bar{B}^*_s\rho$, $\bar{B}_s\rho$, $\bar{B}^*K^*$, and $\bar{B}K^*$ ($\bar{B}^*_s\omega$, $\bar{B}_s\omega$, $\bar{B}^*K^*$, and $\bar{B}K^*$).

In the tensor $J^P=2^+$ case, there are four possible tetraqauark states in total, two around 6291 MeV with $I=1,0$ and two around 6374 MeV with $I=1,0$. The higher $I=1$ ($I=0$) state decays into $\bar{B}^*_s\rho$ and $\bar{B}^*K^*$ ($\bar{B}^*_s\omega$ and $\bar{B}^*K^*$) with partial width ratio around 4.2 (4.1), whereas the $\bar{B}^*K^*$ is the main rearrangement decay mode of the two lower states.

\subsection{The $cn\bar{s}\bar{s}$ and $bn\bar{s}\bar{s}$ systems}

\begin{table}[htbp]
	\caption{Mass spectra for the $cn\bar{s}\bar{s}$ and $bn\bar{s}\bar{s}$ states in units of MeV. The thresholds of $D_sK$ and $\bar{B}_sK$ are used to establish the lower limits for the tetraquark masses in the former and latter cases, respectively. The upper limits are obtained using the effective quark masses: $m_n=361.8$ MeV, $m_s=542.4$ MeV, $m_c=1724.1$ MeV, and $m_b=5054.4$ MeV. We adopt values from the mass splitting model (fourth column) in the discussions.}\label{spectrum-Qnss}\scriptsize
	\begin{tabular}{c|ccccccc}\hline
		\hline\multicolumn{6}{c}{$[cn\bar{s}\bar{s}]^I$ states} \\\hline
		$I(J^{P})$ & $\langle H_{CMI} \rangle$ &Eigenvalue &Mass&Lower limits&Upper limits\\\hline
		$\frac12(2^{+})$ &$\left(\begin{array}{c}95.7\end{array}\right)$&$\left(\begin{array}{c}95.7\end{array}\right)$&$\left(\begin{array}{c}3046.1\end{array}\right)$&$\left(\begin{array}{c}2965.4\end{array}\right)$&$\left(\begin{array}{c}3266.4\end{array}\right)$\\
		$\frac12(1^{+})$ &$\left(\begin{array}{ccc}-39.7&96.0&45.3\\96.0&20.7&-143.7\\45.3&-143.7&-14.7\end{array}\right)$&$\left(\begin{array}{c}158.3\\10.1\\-202.2\end{array}\right)$&$\left(\begin{array}{c}3108.7\\2960.5\\2748.2\end{array}\right)$&$\left(\begin{array}{c}3028.0\\2879.8\\2667.5\end{array}\right)$&$\left(\begin{array}{c}3329.0\\3180.8\\2968.5\end{array}\right)$\\
		$\frac12(0^{+})$ &$\left(\begin{array}{cc}-107.5&248.9\\248.9&42.0\end{array}\right)$&$\left(\begin{array}{c}227.1\\-292.6\end{array}\right)$&$\left(\begin{array}{c}3177.5\\2657.8\end{array}\right)$&$\left(\begin{array}{c}3096.8\\2577.1\end{array}\right)$&$\left(\begin{array}{c}3397.8\\2878.1\end{array}\right)$\\
		\hline\hline
		
	\multicolumn{6}{c}{$[bn\bar{s}\bar{s}]^I$ states} \\\hline
		$I(J^{P})$ & $\langle H_{CM} \rangle$ &Eigenvalue &Mass&Lower limits&Upper limits\\\hline
		$\frac12(2^{+})$ &$\left(\begin{array}{c}76.8\end{array}\right)$&$\left(\begin{array}{c}76.8\end{array}\right)$&$\left(\begin{array}{c}6367.4\end{array}\right)$&$\left(\begin{array}{c}6274.6\end{array}\right)$&$\left(\begin{array}{c}6577.8\end{array}\right)$\\
		$\frac12(1^{+})$ &$\left(\begin{array}{ccc}-35.2&131.2&61.8\\131.2&24.3&-118.8\\61.8&-118.8&6.9\end{array}\right)$&$\left(\begin{array}{c}158.2\\49.3\\-211.5\end{array}\right)$&$\left(\begin{array}{c}6448.7\\6339.9\\6079.1\end{array}\right)$&$\left(\begin{array}{c}6356.0\\6247.1\\5986.3\end{array}\right)$&$\left(\begin{array}{c}6659.2\\6550.3\\6289.5\end{array}\right)$\\
		$\frac12(0^{+})$ &$\left(\begin{array}{cc}-91.2&205.8\\205.8&31.2\end{array}\right)$&$\left(\begin{array}{c}184.7\\-244.7\end{array}\right)$&$\left(\begin{array}{c}6475.2\\6045.9\end{array}\right)$&$\left(\begin{array}{c}6382.5\\5953.1\end{array}\right)$&$\left(\begin{array}{c}6685.7\\6256.3\end{array}\right)$\\
		\hline
	\end{tabular}
\end{table}

\begin{table}[htbp]
	\caption{Rearrangement decays for the $cn\bar{s}\bar{s}$ and $bn\bar{s}\bar{s}$ cases. The two numbers in the parentheses for a decay channel mean dimensionless $100|\mathcal{M}|^2/{\mathcal C}^2$ and dimensional partial width, respectively. The masses and widths are presented in units of MeV.}\label{decay-Qnss}\scriptsize
	\begin{tabular}{c|c|cccc}\hline\hline
		
		$I(J^{P})$&Mass&\multicolumn{3}{c}{Channels}&$\Gamma$\\
		\hline
		\multicolumn{6}{c}{$cn\bar{s}\bar{s}$} \\\hline
		&&$D_s^*K^*$&&\\
		$\frac12(2^{+})$&$\left[\begin{array}{c}3046.1\end{array}\right]$&$\left[\begin{array}{c}(33.3, 119.2)\end{array}\right]$&&&$\left[\begin{array}{c}119.2\end{array}\right]$\\
		&&$D_s^*K^*$&$D_s^*K$&$D_sK^*$&\\
		$\frac12(1^{+})$&$\left[\begin{array}{c}3108.7\\2960.5\\2748.2 \end{array}\right]$&$\left[\begin{array}{c}(47.0, 260.2)\\(1.8, -)\\(1.2,-) \end{array}\right]$&$\left[\begin{array}{c}(0.2, 2.7)\\(1.1, 11.1)\\(40.3, 274.3) \end{array}\right]$&$\left[\begin{array}{c}(6.2, 53.4)\\(31.3, 184.6)\\(4.2, -) \end{array}\right]$&$\left[\begin{array}{c}316.4\\195.7\\274.3\end{array}\right]$\\
		&&$D_s^*K^*$&$D_sK$&&\\
		$\frac12(0^{+})$&$\left[\begin{array}{c}3177.5\\2657.8\end{array}\right]$&$\left[\begin{array}{c}(54.7, 378.1)\\(3.6, -)\end{array}\right]$&$\left[\begin{array}{c}(0.1, 1.1)\\(41.6, 358.1)\end{array}\right]$&&$\left[\begin{array}{c}379.2\\358.1\end{array}\right]$\\
		\hline
		\multicolumn{6}{c}{$bn\bar{s}\bar{s}$} \\\hline
		&&$\bar{B}_s^*K^*$&&\\
		$\frac12(2^{+})$&$\left[\begin{array}{c}6367.4\end{array}\right]$&$\left[\begin{array}{c}(33.3, 36.5)\end{array}\right]$&&&$\left[\begin{array}{c}36.5\end{array}\right]$\\
		&&$\bar{B}_s^*K^*$&$\bar{B}_s^*K$&$\bar{B}_sK^*$&\\
		$\frac12(1^{+})$&$\left[\begin{array}{c}6448.7\\6339.9\\6079.1 \end{array}\right]$&$\left[\begin{array}{c}(42.1, 70.6)\\(6.0, 4.8)\\(2.0, -) \end{array}\right]$&$\left[\begin{array}{c}(0.1, 0.3)\\(0.1, 0.2)\\(41.5, 68.7) \end{array}\right]$&$\left[\begin{array}{c}(12.2, 24.1)\\(27.6, 35.7)\\(1.8, -) \end{array}\right]$&$\left[\begin{array}{c}94.9\\40.6\\68.7\end{array}\right]$\\
		&&$\bar{B}_s^*K^*$&$\bar{B}_sK$&&\\
		$\frac12(0^{+})$&$\left[\begin{array}{c}6475.2\\6045.9\end{array}\right]$&$\left[\begin{array}{c}(54.7, 99.8)\\(3.7, -)\end{array}\right]$&$\left[\begin{array}{c}(0.1, 0.3)\\(41.6, 73.0)\end{array}\right]$&&$\left[\begin{array}{c}100.0\\73.0\end{array}\right]$\\
		\hline\hline
	\end{tabular}
\end{table}

All the relevant states are exotic which cannot mix with the conventional mesons. The predicted masses for the $cn\bar{s}\bar{s}$ ($bn\bar{s}\bar{s}$) states are listed in tTble \ref{spectrum-Qnss}  and their relative positions are shown in Fig.\ref{relative}(c) (Fig.\ref{relative}(i)). The rearrangement decay properties are collected in Table \ref{decay-Qnss}.

In the $cn\bar{s}\bar{s}$ case, our results indicate that all the states lie above the respective thresholds and each state has a dominant decay channel. From Fig.\ref{relative}(c), the masses range from 2658 to 3178 MeV. The lightest state is around 2.66 GeV with $J^P=0^+$, which can be discovered in the $D_sK$ decay mode. Another $0^+$ state mainly decays into $D^*_sK^*$. Its decay into $D_sK$ is suppressed due to a weak coupling amplitude. Of the three $J^P=1^+$ states, only the $D^*_sK$ channel contributes the rearrangement decay of the lightest $\tilde{T}(2748)$. The higher $\tilde{T}(2961)$ dominantly decays into $D_sK^*$ with a branching fraction about 94.3\%. The total width of the highest $\tilde{T}(3109)$ receives contributions mainly from $D_s^*K^*$ and $D_sK^*$ modes. The partial width ratio between these two channels is estimated to be $\Gamma(D_s^*K^*):\Gamma(D_sK^*)\sim4.9$. The $J^P=2^+$ state can be searched for in its unique rearrangement decay channel $D^*_sK^*$.

As for the $bn\bar{s}\bar{s}$ case, from Fig. \ref{relative}(i) and Table \ref{decay-Qnss}, the features are similar to those in the $cn\bar{s}\bar{s}$ case. A slightly different feature lies in the second $1^+$ state. The $\tilde{T}(6340)$ can decay into $\bar{B}^*_sK^*$ but its charmed counterpart $\tilde{T}(2961)$ is below the $D^*_sK^*$ threshold.

\subsection{The $cs\bar{n}\bar{n}$ and $bs\bar{n}\bar{n}$ systems}

\begin{table}[htbp]
	\caption{Mass spectra for the $cs\bar{n}\bar{n}$ and  $bs\bar{n}\bar{n}$ states in units of MeV. The thresholds of $D\bar{K}$ and $\bar{B}\bar{K}$  are used to establish the lower limits for the tetraquark masses in the former and latter cases, respectively.  The upper limits are obtained using the effective quark masses: $m_n=361.8$ MeV, $m_s=542.4$ MeV, $m_c=1724.1$ MeV, and $m_b=5054.4$ MeV. We adopt values from the mass splitting model (fourth column) in the discussions.}\label{spectrum-Qsnn}
	\scriptsize
	\begin{tabular}{c|ccccccc}\hline
		\hline\multicolumn{6}{c}{$cs\bar{n}\bar{n}$ states} \\\hline
		$I(J^{P})$ & $\langle H_{CM} \rangle$ &Eigenvalue &Mass&Lower limits&Upper limits\\\hline
		$1(2^{+})$ &$\left(\begin{array}{c}127.7\end{array}\right)$&$\left(\begin{array}{c}127.7\end{array}\right)$&$\left(\begin{array}{c}2987.5\end{array}\right)$&$\left(\begin{array}{c}2892.3\end{array}\right)$&$\left(\begin{array}{c}3117.8\end{array}\right)$\\
		$1(1^{+})$ &$\left(\begin{array}{ccc}-7.2&96.8&45.6\\96.8&67.5&-143.1\\45.6&-143.1&14.4\end{array}\right)$&$\left(\begin{array}{c}198.2\\42.5\\-166.1\end{array}\right)$&$\left(\begin{array}{c}3058.0\\2902.3\\2693.7\end{array}\right)$&$\left(\begin{array}{c}2962.8\\2807.1\\2598.5\end{array}\right)$&$\left(\begin{array}{c}3188.3\\3032.6\\2824.0\end{array}\right)$\\
		$1(0^{+})$ &$\left(\begin{array}{cc}-74.7&247.9\\247.9&90.4\end{array}\right)$&$\left(\begin{array}{c}269.1\\-253.4\end{array}\right)$&$\left(\begin{array}{c}3128.9\\2606.4\end{array}\right)$&$\left(\begin{array}{c}3033.7\\2511.2\end{array}\right)$&$\left(\begin{array}{c}3259.2\\2736.7\end{array}\right)$\\
		$0(2^{+})$ &$\left(\begin{array}{c}138.5\end{array}\right)$&$\left(\begin{array}{c}138.5\end{array}\right)$&$\left(\begin{array}{c}2998.3\end{array}\right)$&$\left(\begin{array}{c}2903.1\end{array}\right)$&$\left(\begin{array}{c}3128.6\end{array}\right)$\\
		$0(1^{+})$ &$\left(\begin{array}{ccc}-198.8&96.8&114.1\\96.8&-134.9&-143.1\\114.1&-143.1&-7.2\end{array}\right)$&$\left(\begin{array}{c}93.7\\-75.5\\-359.2\end{array}\right)$&$\left(\begin{array}{c}2953.5\\2784.3\\2500.6\end{array}\right)$&$\left(\begin{array}{c}2858.3\\2689.1\\2405.4\end{array}\right)$&$\left(\begin{array}{c}3083.8\\2914.6\\2630.9\end{array}\right)$\\
		$0(0^{+})$ &$\left(\begin{array}{cc}-367.5&247.9\\247.9&-180.8\end{array}\right)$&$\left(\begin{array}{c}-9.3\\-539.0\end{array}\right)$&$\left(\begin{array}{c}2850.5\\2320.7\end{array}\right)$&$\left(\begin{array}{c}2755.3\\2225.6\end{array}\right)$&$\left(\begin{array}{c}2980.8\\2451.1\end{array}\right)$\\
		
		\hline\hline\multicolumn{6}{c}{$bs\bar{n}\bar{n}$ states} \\\hline
		$I(J^{P})$ & $\langle H_{CM} \rangle$ &Eigenvalue  &Mass&Lower limits&Upper limits\\\hline
		$1(2^{+})$ &$\left(\begin{array}{c}107.7\end{array}\right)$&$\left(\begin{array}{c}107.7\end{array}\right)$&$\left(\begin{array}{c}6307.7\end{array}\right)$&$\left(\begin{array}{c}6214.9\end{array}\right)$&$\left(\begin{array}{c}6428.1\end{array}\right)$\\
		$1(1^{+})$ &$\left(\begin{array}{ccc}-3.2&132.8&62.6\\132.8&71.5&-117.7\\62.6&-117.7&38.4\end{array}\right)$&$\left(\begin{array}{c}199.8\\82.2\\-175.3\end{array}\right)$&$\left(\begin{array}{c}6399.7\\6282.2\\6024.7\end{array}\right)$&$\left(\begin{array}{c}6307.0\\6189.4\\5931.9\end{array}\right)$&$\left(\begin{array}{c}6520.2\\6402.6\\6145.1\end{array}\right)$\\
		$1(0^{+})$ &$\left(\begin{array}{cc}-58.7&203.8\\203.8&78.4\end{array}\right)$&$\left(\begin{array}{c}224.9\\-205.1\end{array}\right)$&$\left(\begin{array}{c}6424.8\\5994.8\end{array}\right)$&$\left(\begin{array}{c}6332.1\\5902.1\end{array}\right)$&$\left(\begin{array}{c}6545.3\\6115.3\end{array}\right)$\\
		$0(2^{+})$ &$\left(\begin{array}{c}112.5\end{array}\right)$&$\left(\begin{array}{c}112.5\end{array}\right)$&$\left(\begin{array}{c}6312.5\end{array}\right)$&$\left(\begin{array}{c}6219.7\end{array}\right)$&$\left(\begin{array}{c}6432.9\end{array}\right)$\\
		$0(1^{+})$ &$\left(\begin{array}{ccc}-164.8&132.8&156.5\\132.8&-142.9&-117.7\\156.5&-117.7&-19.2\end{array}\right)$&$\left(\begin{array}{c}86.1\\-23.2\\-389.8\end{array}\right)$&$\left(\begin{array}{c}6286.0\\6176.8\\5810.1\end{array}\right)$&$\left(\begin{array}{c}6193.3\\6084.0\\5717.4\end{array}\right)$&$\left(\begin{array}{c}6406.5\\6297.2\\5930.6\end{array}\right)$\\
		$0(0^{+})$ &$\left(\begin{array}{cc}-303.5&203.8\\203.8&-156.8\end{array}\right)$&$\left(\begin{array}{c}-13.5\\-446.7\end{array}\right)$&$\left(\begin{array}{c}6186.4\\5753.2\end{array}\right)$&$\left(\begin{array}{c}6093.7\\5660.5\end{array}\right)$&$\left(\begin{array}{c}6306.9\\5873.7\end{array}\right)$\\ \hline
			\end{tabular}
\end{table}

\setlength{\tabcolsep}{0.001mm}
\begin{table}[htbp]\scriptsize
	\caption{Rearrangement decays for the $cs\bar{n}\bar{n}$ and $bs\bar{n}\bar{n}$ cases. The two numbers in the parentheses for a decay channel mean dimensionless $100|\mathcal{M}|^2/{\mathcal C}^2$ and dimensional partial width, respectively. The masses and widths are presented in units of MeV.}\label{decay-Qsnn}
	\begin{tabular}{c|c|cccc}\hline\hline
			
		$I(J^{P})$&Mass&\multicolumn{3}{c}{Channels}&$\Gamma$\\ \hline
		\multicolumn{6}{c}{$cs\bar{n}\bar{n}$} \\\hline
		&&$D^*\bar{K}^*$&&\\
		$1(2^{+})$&$\left[\begin{array}{c}2987.5\end{array}\right]$&$\left[\begin{array}{c}(33.3, 180.1)\end{array}\right]$&&&$\left[\begin{array}{c}180.1\end{array}\right]$
	\\
		&&$D^*\bar{K}^* $&$D^*\bar{K}$&$D\bar{K}^*$&\\
		$1(1^{+})$&$\left[\begin{array}{c}3058.0\\2902.3\\2693.7 \end{array}\right]$&$\left[\begin{array}{c}(46.9, 329.8)\\(2.2, 0.3)\\(0.9, -) \end{array}\right]$&$\left[\begin{array}{c}(0.4, 4.5)\\(1.1, 12.1)\\(40.2, 333.5) \end{array}\right]$&$\left[\begin{array}{c}(7.1, 69.0)\\(30.7, 225.3)\\(3.9, -) \end{array}\right]$&$\left[\begin{array}{c}403.3\\237.7\\333.5\end{array}\right]$\\
		&&$D^*\bar{K}^*$&$D\bar{K}$&&\\
		$1(0^{+})$&$\left[\begin{array}{c}3128.9\\2606.4\end{array}\right]$&$\left[\begin{array}{c}(55.1, 451.1)\\(3.2,  -)\end{array}\right]$&$\left[\begin{array}{c}(0.1, 2.1)\\(41.5, 419.3)\end{array}\right]$&&$\left[\begin{array}{c}453.2\\419.3\end{array}\right]$\\
		\hline
&&$D^*\bar{K}^*$&&\\

$0(2^{+})$&$\left[\begin{array}{c}2998.3\end{array}\right]$&$\left[\begin{array}{c}(66.7, 380.2)\end{array}\right]$&&&$\left[\begin{array}{c}380.2\end{array}\right]$\\

&&$D^*\bar{K}^* $&$D^*\bar{K}$&$D\bar{K}^*$&\\

$0(1^{+})$&$\left[\begin{array}{c}2953.5\\2784.3\\2500.6 \end{array}\right]$&$\left[\begin{array}{c}(47.6, 203.0)\\(2.4,-)\\(0.1, -) \end{array}\right]$&$\left[\begin{array}{c}(1.3, 14.5)\\(1.8, 17.8)\\(55.2,  -) \end{array}\right]$&$\left[\begin{array}{c}(10.6, 88.5)\\(47.6, 151.9)\\(0.1, -) \end{array}\right]$&$\left[\begin{array}{c}306.1\\169.7\\-\end{array}\right]$\\

&&$D^*\bar{K}^*$&$D\bar{K}$&&\\

$0(0^{+})$&$\left[\begin{array}{c}2850.5\\2320.7\end{array}\right]$&$\left[\begin{array}{c}(41.4, -)\\(0.3, -)\end{array}\right]$&$\left[\begin{array}{c}(2.7, 34.4)\\(55.6, - )\end{array}\right]$&&$\left[\begin{array}{c}34.4\\-\end{array}\right]$\\\hline
		
	\multicolumn{6}{c}{$bs\bar{n}\bar{n}$} \\\hline
		&&$\bar{B}^*\bar{K}^*$&&&\\
		$1(2^{+})$&$\left[\begin{array}{c}6307.7\end{array}\right]$&$\left[\begin{array}{c}(33.3, 46.3)\end{array}\right]$&&&$\left[\begin{array}{c}46.3\end{array}\right]$\\
		&&$\bar{B}^*\bar{K}^* $&$\bar{B}^*\bar{K}$&$\bar{B}\bar{K}^*$&\\
		$1(1^{+})$&$\left[\begin{array}{c}6399.7\\6282.2\\6024.7 \end{array}\right]$&$\left[\begin{array}{c}(41.6, 81.4)\\(6.7, 7.9)\\(1.7, -) \end{array}\right]$&$\left[\begin{array}{c}(0.2, 0.6)\\(0.1, 0.2)\\(41.4, 77.9) \end{array}\right]$&$\left[\begin{array}{c}(13.3, 29.3)\\(26.8, 41.5)\\(1.6, -) \end{array}\right]$&$\left[\begin{array}{c}111.3\\49.6\\77.9\end{array}\right]$\\
		
		&&$\bar{B}^*\bar{K}^*$&$\bar{B}\bar{K}$&&\\
		$1(0^{+})$&$\left[\begin{array}{c}6424.8\\5994.8\end{array}\right]$&$\left[\begin{array}{c}(55.2, 114.7)\\(3.2,  -)\end{array}\right]$&$\left[\begin{array}{c}(0.2, 0.5)\\(41.5, 82.1)\end{array}\right]$&&$\left[\begin{array}{c}115.2\\82.1\end{array}\right]$\\\hline
		
&&$\bar{B}^*\bar{K}^*$&&&\\

$0(2^{+})$&$\left[\begin{array}{c}6312.5\end{array}\right]$&$\left[\begin{array}{c}(66.7, 95.0)\end{array}\right]$&&&$\left[\begin{array}{c}95.0\end{array}\right]$\\

&&$\bar{B}^*\bar{K}^*$&$\bar{B}^*\bar{K}$&$\bar{B}\bar{K}^*$&\\

$0(1^{+})$&$\left[\begin{array}{c}6286.0\\6176.8\\5810.1 \end{array}\right]$&$\left[\begin{array}{c}(34.1, 41.3)\\(15.8, -)\\(0.1, -) \end{array}\right]$&$\left[\begin{array}{c}(0.2, 0.6)\\(2.9, 7.2)\\(55.2, -) \end{array}\right]$&$\left[\begin{array}{c}(31.2, 49.2)\\(27.1, 7.8)\\(0.1, -) \end{array}\right]$&$\left[\begin{array}{c}91.0\\14.9\\-\end{array}\right]$\\

&&$\bar{B}^*\bar{K}^*$&$\bar{B}\bar{K}$&&\\

$0(0^{+})$&$\left[\begin{array}{c}6186.4\\5753.2\end{array}\right]$&$\left[\begin{array}{c}(41.5, -)\\(0.2,-)\end{array}\right]$&$\left[\begin{array}{c}(2.9, 7.8)\\(55.4,  -)\end{array}\right]$&&$\left[\begin{array}{c}7.8\\-\end{array}\right]$\\

\hline
	\end{tabular}
\end{table}

The  tetraquark ground  states in these two systems are also flavor exotic. Each system comprises six states with $I=1$ and six with $I=0$. We present the estimated masses in Table \ref{spectrum-Qsnn} and show the relative positions in Fig. \ref{relative}(d) and Fig. \ref{relative}(j). Table \ref{decay-Qsnn} gives the rearrangement decay properties.

We first discuss the $I=0$ $cs\bar{n}\bar{n}$ case. Our results show that the estimated mass of the higher $J^P=0^+$ state is 2850.5 MeV, which is in good agreement with the measured mass of the observed  $T_{cs0}^*(2900)^0$  \cite{LHCb:2020bls,LHCb:2020pxc}. The calculated width of $\tilde{T}^f(2851)$ is 34.4 MeV. This value is larger than our previous result \cite{Cheng:2020nho}, although it is still smaller than the measured with of the  $T_{cs0}^*(2900)^0$ , $57\pm12\pm4$ MeV, Note that our estimation for the width only involves the rearrangement channel $D\bar{K}$ and the model is very crude. It is acceptable to assign the  $T_{cs0}^*(2900)^0$  as the higher $J^P=0^+$ $cs\bar{n}\bar{n}$ state. We this assignment, we may consistently understand the structures of both  $T_{cs0}^*(2900)^0$  and $T^s_{c\bar{s}0}(2900)$. Besides this $0^+$ tetraquark state around 2900 MeV, one should also note the other $0^+$ state $\tilde{T}(2321)$. It is expected to be stable with respect to the strong interaction. Searching for this state in the $D^+\pi^-\pi^0$ invariant mass is called for.

From Fig. \ref{relative}(d) and Table \ref{decay-Qsnn}, the lightest $I(J^P)=0(1^+)$ state $\tilde{T}^f(2501)$ is around the $D^*\bar{K}$ threshold and may be stable or narrow. It could be searched for in the $D^0\pi^+K^-$ channel in experiments. 
The higher $0(1^+)$ state $\tilde{T}^f(2784)$ dominantly decays into $D^*\bar{K}$ and $D\bar{K}^*$. Their partial width ratio is predicted to be
\begin{eqnarray}
	\Gamma(D^*\bar{K}):\Gamma(D\bar{K}^*)\simeq 0.12.
\end{eqnarray}
The decay of the highest $0(1^+)$ state $\tilde{T}^f(2954)$ receives contributions mainly from the $D^*\bar{K}^*$ and $D\bar{K}^*$ channels, and subdominantly from the $D^*\bar{K}$. Although the phase space of the channel $D^*\bar{K}$ is the largest one, the weak coupling leads to a small width. In our model, the partial width ratio is found to be
\begin{eqnarray}
	\Gamma(D^*\bar{K}^*):\Gamma(D^*\bar{K}):\Gamma(D\bar{K}^*)\simeq 14.0:1.0:6.1.
\end{eqnarray}
 The last of the six $I=0$ states is the broad $\tilde{T}^f(2998)$ with $J^P=2^+$, which has only one rearrangement decay channel $D^*\bar{K}^*$.

From Fig. \ref{relative}(d), the $I=1$ $cs\bar{n}\bar{n}$ states have generally lager masses than the $I=0$ $cs\bar{n}\bar{n}$ states. Since the rearrangement decay patterns in these two cases are the same, it is evident that all the $I=1$ states are unstable and have broad widths. For the $I=1$ $cs\bar{n}\bar{n}$ case, the partial decay ratios can be derived with data in Table \ref{decay-Qsnn}.
 
The higher $I(J^P)=1(0^+)$ state $\tilde{T}^a(3129)$ is the heaviest and broadest $cs\bar{n}\bar{n}$ tetraquark. Its mass is much larger than the $D\bar{K}$ threshold, but the state dominantly decays into $D^*\bar{K}^*$ because of the strong coupling amplitude. Due to the combined effects of phase space and coupling, the lower $1(0^+)$ $\tilde{T}^a(2606)$ has a width comparable to that of the higher $\tilde{T}^a(3129)$.
Each of the three $I(J^P)=1(1^+)$ states has a dominant rearrangement decay mode. They are $D^*\bar{K}$, $D\bar{K}^*$, and $D^*\bar{K}^*$ for the lightest $\tilde{T}^a(2694)$, the higher $\tilde{T}^a(2902)$, and the heaviest $\tilde{T}^a(3058)$, respectively. The branching fraction for the $\tilde{T}^a(2902)$ into $D\bar{K}^*$ can reach up to $95\%$ of the total rearrangement decay width. The $\tilde{T}^a(3058)$ also decays into $D\bar{K}^*$ significantly, with the ratio of partial widths $\Gamma(D^*\bar{K}^*):\Gamma(D\bar{K}^*)\sim 4.8$. The width of the tensor state $\tilde{T}^a(2988)$ is smaller than its $I=0$ counterpart, which results from its slightly lower mass and weaker coupling with the $D^*\bar{K}^*$ channel.\\

We now move on to the $b$ quark sector and study the $bs\bar{n}\bar{n}$  tetraquark ground  states. In Ref. \cite{Yu:2017pmn}, the author proposed to search for a $bs\bar{u}\bar{d}$ tetraquark state through its weak decay, provided that such a state exists below the $\bar{B}\bar{K}$ threshold.
In Ref. \cite{Lu:2020qmp}, the authors adopted an extended relativized quark model to study the mass of open bottom tetraquark states. Their results showed that all of the predicted $nn\bar{s}\bar{b}$ masses lie above the respective thresholds. This means that the rearrangement decay can occur for all the $bs\bar{n}\bar{n}$ tetraquark states. 
 From Table \ref{decay-Qsnn}, one observes that the features for the $bs\bar{n}\bar{n}$ case are similar to those for the $cs\bar{n}\bar{n}$ case. Assuming that the decay parameter ${\cal C}$ takes a similar value as in the $c$ quark sector, one finds that the widths of the $bs\bar{n}\bar{n}$ states are smaller. 
 
For the $I=0$ case, the lighter $0^+$ state with mass $5753.2$ MeV and the lightest $1^+$ state with mass $5810.1$ MeV are found to be stable, in agreement with our previous results \cite{Cheng:2020nho}. The experiments could choose the $B^-\pi^-\pi^0$ channel to detect the lighter $0^+$ state. Searching for the lightest $1^+$ state in the $\bar{B}\bar{K}\gamma$ channels is highly desirable.  The higher $0^+$ state is a narrow $bs\bar{n}\bar{n}$ with a width of several MeV, decaying exclusively into $\bar{B}\bar{K}$.
For the higher $1^+$ state $\tilde{T}^f(6177)$, the two dominant decay modes  $\bar{B}^*\bar{K}$ and $\bar{B}\bar{K}^*$ have similar partial widths, as the stronger coupling to the latter channel compensates for its smaller phase space. 
For the highest $1^+$ state around $6286$ MeV, the dominant rearrangement decay modes are  $\bar{B^*}\bar{K^*}$ and $\bar{B}\bar{K^*}$, which have comparable coupling strengths and partial widths. Regarding the remaining $2^+$ state, the width comes mainly from the $\bar{B}^*\bar{K}^*$ channel.

The differentiating features between the $I=1$ and $I=0$ $cs\bar{n}\bar{n}$ states are similarly observed in the $bs\bar{n}\bar{n}$ cases. These features help to understand the properties of the $I=1$ $bs\bar{n}\bar{n}$ tetraquarks.


 \subsection{The $cs\bar{s}\bar{n}$ and $bs\bar{s}\bar{n}$ systems}

 \begin{table}[htbp]
	\caption{Mass spectra for the $cs\bar{s}\bar{n}$ and $bs\bar{s}\bar{n}$ states in units of MeV. The thresholds of $D_s\bar{K}$ and $\bar{B}_s\bar{K}$ are used to establish the lower limits for the tetraquark masses in the former and latter cases, respectively.   The upper limits are obtained using the effective quark masses: $m_n=361.8$ MeV, $m_s=542.4$ MeV, $m_c=1724.1$ MeV, and $m_b=5054.4$ MeV. We adopt values from the mass splitting model (fourth column) in the discussions.}\label{spectrum-Qssn}
	\scriptsize
	\begin{tabular}{c|ccccccc}\hline
       \hline\multicolumn{6}{c}{$cs\bar{s}\bar{n}$ states} \\\hline
		$I(J^{P})$ & $\langle H_{CM} \rangle$ &Eigenvalue  &Mass&Lower limits&Upper limits\\\hline
		$\frac12(2^{+})$ &$\left(\begin{array}{cc}117.5&-25.5\\-25.5&99.5\end{array}\right)$&$\left(\begin{array}{c}135.5\\81.5\end{array}\right)$&$\left(\begin{array}{c}3085.8\\3031.8\end{array}\right)$&$\left(\begin{array}{c}3005.2\\2951.2\end{array}\right)$&$\left(\begin{array}{c}3306.2\\3252.2\end{array}\right)$\\
		$\frac12(1^{+})$ &$\left(\begin{array}{cccccc}-161.2&25.5&-41.5&60.8&71.7&-35.2\\25.5&-12.0&60.8&-16.6&-35.2&28.7\\-41.5&60.8&42.7&0&30.0&-118.2\\60.8&-16.6&0&-85.3&-118.2&12.0\\71.7&-35.2&30.0&-118.2&1.1&0\\-35.2&28.7&-118.2&12.0&0&-2.1\end{array}\right)$&$\left(\begin{array}{c}152.2\\85.9\\13.0\\-66.6\\-115.6\\-285.8\end{array}\right)$&$\left(\begin{array}{c}3102.5\\3036.2\\2963.4\\2883.7\\2834.8\\2664.6\end{array}\right)$&$\left(\begin{array}{c}3021.9\\2955.6\\2882.7\\2803.1\\2754.1\\2583.9\end{array}\right)$&$\left(\begin{array}{c}3322.9\\3256.6\\3183.7\\3104.1\\3055.1\\2884.9\end{array}\right)$\\
		$\frac12(0^{+})$ &$\left(\begin{array}{cccc}-300.5&50.9&-52.0&204.8\\50.9&-67.7&204.8&-20.8\\-52.0&204.8&65.6&0\\204.8&-20.8&0&-131.2\end{array}\right)$&$\left(\begin{array}{c}215.7\\7.5\\-199.7\\-457.3\end{array}\right)$&$\left(\begin{array}{c}3166.0\\2957.8\\2750.7\\2493.0\end{array}\right)$&$\left(\begin{array}{c}3085.4\\2877.2\\2670.0\\2412.4\end{array}\right)$&$\left(\begin{array}{c}3386.4\\3178.2\\2971.0\\2713.4\end{array}\right)$\\
		\hline\hline
		\multicolumn{6}{c}{$bs\bar{s}\bar{n}$ states} \\\hline
		$I(J^{P})$ & $\langle H_{CM} \rangle$ &Eigenvalue  &Mass&Lower limits&Upper limits\\\hline
		$\frac12(2^{+})$ &$\left(\begin{array}{cc}91.8&25.7\\25.7&79.6\end{array}\right)$&$\left(\begin{array}{c}112.2\\59.2\end{array}\right)$&$\left(\begin{array}{c}6402.7\\6349.8\end{array}\right)$&$\left(\begin{array}{c}6164.4\\6111.4\end{array}\right)$&$\left(\begin{array}{c}6613.2\\6560.2\end{array}\right)$\\
		$\frac12(1^{+})$ &$\left(\begin{array}{cccccc}-127.5&-25.7&41.0&96.4&113.6&34.8\\-25.7&-8.1&96.4&16.4&34.8&45.4\\41.0&96.4&46.7&0&-30.3&-93.1\\96.4&16.4&0&-93.3&-93.1&-12.1\\113.6&34.8&-30.3&-93.1&-10.9&0\\34.8&45.4&-93.1&-12.1&0&21.9\end{array}\right)$&$\left(\begin{array}{c}148.0\\82.6\\37.7\\-9.8\\-117.1\\-312.8\end{array}\right)$&$\left(\begin{array}{c}6438.5\\6373.1\\6328.2\\6280.8\\6173.5\\5977.8\end{array}\right)$&$\left(\begin{array}{c}6200.2\\6134.8\\6089.9\\6042.4\\5935.1\\5739.4\end{array}\right)$&$\left(\begin{array}{c}6649.0\\6583.6\\6538.7\\6491.2\\6383.9\\6188.2\end{array}\right)$\\
		$\frac12(0^{+})$ &$\left(\begin{array}{cccc}-237.2&-51.5&52.5&161.2\\-51.5&-52.0&161.2&21.0\\52.5&161.2&53.6&0\\161.2&21.0&0&-107.2\end{array}\right)$&$\left(\begin{array}{c}172.2\\4.2\\-148.9\\-370.3\end{array}\right)$&$\left(\begin{array}{c}6462.7\\6294.8\\6141.7\\5920.3\end{array}\right)$&$\left(\begin{array}{c}6224.4\\6056.4\\5903.3\\5681.9\end{array}\right)$&$\left(\begin{array}{c}6673.2\\6505.2\\6352.1\\6130.7\end{array}\right)$\\\hline
			\end{tabular}
\end{table}

 \begin{table}[htbp]\scriptsize
 	\caption{Rearrangement decays for the $cs\bar{s}\bar{n}$ and $bs\bar{s}\bar{n}$ cases. The two numbers in the parentheses for a decay channel mean dimensionless $100|\mathcal{M}|^2/{\mathcal C}^2$ and dimensional partial width, respectively. The masses and widths are presented in units of MeV.}\label{decay-Qssn}
 	\begin{tabular}{c|c|ccccccc|c}\hline\hline
 		$I(J^{P})$&Mass&\multicolumn{7}{c}{Channels}&$\Gamma$\\	\hline
 		\multicolumn{10}{c}{$cs\bar{s}\bar{n}$} \\\hline

 		&&$D^*_s\bar{K}^*$&$D^*\phi$&&&&&&\\
 		$\frac12(2^{+})$&$\left[\begin{array}{c}3085.8\\3031.8\end{array}\right]$&$\left[\begin{array}{c}(100.0, 247.0)\\(0, 0)\end{array}\right]$&$\left[\begin{array}{c}(11.1, 24.1)\\(88.9, 50.9)\end{array}\right]$&&&&&&$\left[\begin{array}{c}271.1\\50.9\end{array}\right]$\\
 		&&$D_s^*\bar{K}^*$&$D_s^*\bar{K}$&$D_s\bar{K}^*$&$D^*\phi$&$D^*\eta$&$D^*\eta^{\prime}$&$D\phi$&\\
 		$\frac12(1^{+})$&$\left[\begin{array}{c}3102.5\\3036.2\\2963.4\\2883.7\\2834.8\\2664.6\end{array}\right]$&$\left[\begin{array}{c}(76.1, 204.9)\\(23.2, 36.3)\\(0.2, -)\\(0.5, -)\\(0.0, -)\\(0.0, -)\end{array}\right]$&$\left[\begin{array}{c}(0.1, 0.6)\\(0.5, 2.6)\\(2.0, 10.1)\\(0.9, 3.9)\\(7.0, 29.3)\\(89.4, 200.8)\end{array}\right]$&$\left[\begin{array}{c}(2.4, 10.2)\\(18.2, 68.5)\\(33.2, 99.2)\\(46.0, 66.5)\\(0.3,-)\\(0.0, )\end{array}\right]$&$\left[\begin{array}{c}(21.7, 52.9)\\(66.8, 56.0)\\(9.3,-)\\(0.0,-)\\(1.4, -)\\(0.9, -)\end{array}\right]$&$\left[\begin{array}{c}(0.9, 5.4)\\(1.5, 8.6)\\(0.8, 4.6)\\(11.7, 59.4)\\(23.2, 111.4)\\(8.9, 28.8)\end{array}\right]$&$\left[\begin{array}{c}(1.0, 3.3)\\(1.7, 4.1)\\(0.9,-)\\(13.2,- )\\(26.1,-)\\(10.0, -)\end{array}\right]$&$\left[\begin{array}{c}(6.6, 27.7)\\(0.5, 1.7)\\(22.3, 59.9)\\(37.9, -)\\(30.4, -)\\(2.2, -)\end{array}\right]$&$\left[\begin{array}{c}305.1\\177.8\\173.7 \\129.8 \\140.7 \\229.7 \end{array}\right]$\\
 		&&$D_s^*\bar{K}^*$&$D_s\bar{K}$&$D^*\phi$&$D\eta$&$D\eta^{\prime}$&&&\\
 		$\frac12(0^{+})$&$\left[\begin{array}{c}3166.0\\2957.8\\2750.7\\2493.0\end{array}\right]$&$\left[\begin{array}{c}(62.8, 211.0)\\(35.8,-)\\(1.3,-)\\(0.0, -)\end{array}\right]$&$\left[\begin{array}{c}(0.1, 0.4)\\(0.8, 4.9)\\(18.1, 90.8)\\(81.0, 146.3)\end{array}\right]$&$\left[\begin{array}{c}(47.3, 152.1)\\(45.6, -)\\(5.6,-)\\(1.5, -)\end{array}\right]$&$\left[\begin{array}{c}(0.2, 1.0)\\(2.7, 17.3)\\(30.4, 170.1)\\(13.9, 43.0)\end{array}\right]$&$\left[\begin{array}{c}(0.2, 0.9)\\(3.0, 10.6)\\(34.1, -)\\(15.6,-)\end{array}\right]$&&&$\left[\begin{array}{c}365.4\\32.7 \\260.9 \\189.2 \end{array}\right]$\\
 		\hline
 		\multicolumn{10}{c}{$bs\bar{s}\bar{n}$ } \\\hline
 		&&$\bar{B}^*_sK^*$&$\bar{B}^*\phi$&&&&&&\\
 		$\frac12(2^{+})$&$\left[\begin{array}{c}6402.7\\6349.8\end{array}\right]$&$\left[\begin{array}{c}(99.7, 68.9)\\(0.3, 0.1)\end{array}\right]$&$\left[\begin{array}{c}(11.1, 4.6)\\(92.0, 16.5)\end{array}\right]$&&&&&&$\left[\begin{array}{c}73.5\\16.7\end{array}\right]$\\
 		&&$\bar{B}_s^*\bar{K}^*$&$\bar{B}_s^*\bar{K}$&$\bar{B}_s\bar{K}^*$&$\bar{B}^*\phi$&$\bar{B}^*\eta$&$\bar{B}^*\eta^{\prime}$&$\bar{B}\phi$&\\
 		$\frac12(1^{+})$&$\left[\begin{array}{c}6438.5\\6373.1\\6328.2\\6280.8\\6173.5\\5977.8\end{array}\right]$&$\left[\begin{array}{c}(59.6, 48.2)\\(28.4, 16.3)\\(3.7, 1.2)\\(8.2, -)\\(0.1, -)\\(0.0, -)\end{array}\right]$&$\left[\begin{array}{c}(0.0, 0.1)\\(0.0, 0.0)\\(0.3, 0.4)\\(0.2, 0.2)\\(12.2, 12.6)\\(87.2, 45.3)\end{array}\right]$&$\left[\begin{array}{c}(9.2, 8.8)\\(61.4, 47.1)\\(3.4, 2.0)\\(25.8, 8.5)\\(0.2,-)\\(0, -)\end{array}\right]$&$\left[\begin{array}{c}(28.8, 20.8)\\(16.9, 6.8)\\(41.9, -)\\(7.4,-)\\(3.9, -)\\(1.1, -)\end{array}\right]$&$\left[\begin{array}{c}(0.3, 0.5)\\(0.3, 0.4)\\(0.2, 0.2)\\(4.7, 6.3)\\(31.4, 36.5)\\(10.3, 7.1)\end{array}\right]$&$\left[\begin{array}{c}(0.4, 0.3)\\(0.3, 0.2)\\(0.2, 0.1)\\(5.2, -)\\(35.3, -)\\(11.5, -)\end{array}\right]$&$\left[\begin{array}{c}(11.9, 10.5)\\(0.5, 0.3)\\(36.9, 15.2)\\(44.6,-)\\(5.2, -)\\(0.9, -)\end{array}\right]$&$\left[\begin{array}{c}89.2\\71.2\\19.2 \\15.0 \\49.1 \\52.4 \end{array}\right]$\\
 		&&$\bar{B}_s^*\bar{K}^*$&$\bar{B}_s\bar{K}$&$\bar{B}^*\phi$&$\bar{B}\eta$&$\bar{B}\eta^{\prime}$&&&\\
 		$\frac12(0^{+})$&$\left[\begin{array}{c}6462.7\\6294.8\\6141.7\\5920.3\end{array}\right]$&$\left[\begin{array}{c}(64.9, 57.0)\\(34.4, -)\\(0.7, -)\\(0.0, -)\end{array}\right]$&$\left[\begin{array}{c}(0.0, 0.1)\\(0.6, 0.8)\\(14.4, 15.6)\\(85.0, 41.7)\end{array}\right]$&$\left[\begin{array}{c}(45.5, 36.7)\\(46.0, -)\\(6.7, -)\\(1.8, $-$)\end{array}\right]$&$\left[\begin{array}{c}(0.2, 0.3)\\(3.6, 5.3)\\(31.6, 38.0)\\(11.6, 7.6)\end{array}\right]$&$\left[\begin{array}{c}(0.2, 0.2)\\(4.1, 2.3)\\(35.5, -)\\(13.1, -)\end{array}\right]$&&&$\left[\begin{array}{c}94.3\\8.4 \\53.6 \\49.3 \end{array}\right]$\\
 		\hline\hline
 	\end{tabular}
 \end{table}

Like the $I=1/2$ $Qn\bar{n}\bar{n}$ cases, the $Qs\bar{s}\bar{n}$ states resemble excited $Q\bar{n}$ mesons. The mixing between these three configurations cannot be ruled out, but we defer such discussions to future work. In both the $cs\bar{s}\bar{n}$ and $bs\bar{s}\bar{n}$ systems, there are twelve tetraquark states. The masses of all these states can be found in Table \ref{spectrum-Qssn}, Fig. \ref{relative}(e), and Fig. \ref{relative}(k). Table \ref{decay-Qssn} lists their rearrangement decay properties.

All the $cs\bar{s}\bar{n}$ states lie above their respective rearrangement decay thresholds, indicating that they can fall apart into two hadrons. The spin of the lightest state $\tilde{T}(2493)$ is $J=0$. It mainly decays into $D_s\bar{K}$ and $D\eta$. Our calculation yields the partial width ratio of
 \begin{eqnarray}
	\Gamma(D_s\bar{K}):\Gamma(D\eta)\simeq 3.4.
\end{eqnarray}
The other three $J^P=0^+$ tetraquark states may also be searched for in the $D\eta$ or $D_s\bar{K}$ channel. Note that $\tilde{T}(2958)$, the third lightest among the four $0^+$ states, is the narrowest tetraquark. For the six $J^P=1^+$ $cs\bar{s}\bar{n}$ states, both decay channels $D_s^*\bar{K}$ and $D^*\eta$ are allowed. The partial width ratios for each state can be readily derived from table \ref{decay-Qssn}. The ratios of the partial widths for the decay channel $D^*\eta$ among the six states can be predicted as 
  \begin{align}
 	&\Gamma(\tilde{T}(3103)\to D^*\eta):\Gamma(\tilde{T}(3036)\to D^*\eta):\Gamma(\tilde{T}(2963)\to D^*\eta):\Gamma(\tilde{T}(2884)\to D^*\eta):\nonumber\\
 	&\Gamma(\tilde{T}(2834)\to D^*\eta):
 	\Gamma(\tilde{T}(2665)\to D^*\eta)\sim 1.2:1.9:1.0:12.9:24.2:6.3.
 \end{align}
The corresponding ratios  for the $D_s^*\bar{K}$ channel are 
  \begin{align}
1.0:4.3:16.8:6.5:48.8:334.7.
  \end{align} These values would provide information to identify possible structures, once all the six states are observed. The two $J^P=2^+$ tetraquarks have different dominant rearrangement decay channels. The higher state $\tilde{T}(3086)$ mainly decays into $D^*_s\bar{K^*}$ with a branching ratio of about 91\%, while the lighter $\tilde{T}(3032)$ mainly decays into $D^*\phi$ with a branching ratio of nearly 100\%.\\

For the $bs\bar{s}\bar{n}$ system, the features in spectrum and rearrangement decay are similar to the $cs\bar{s}\bar{n}$ case. From Table \ref{decay-Qssn}, both the lowest and the second lowest $J^P=0^+$ states have two dominant rearrangement decay channels, $\bar{B}_s\bar{K}$ and $\bar{B}\eta$. The third lowest $0^+$ state is again the narrowest  tetraquark. It mainly decays into $\bar{B}\eta$ and $\bar{B}\eta^{\prime}$, while the channel $\bar{B}_s\bar{K}$ is relatively suppressed. For the highest $0^+$ state, the channels $\bar{B}_s^*\bar{K}^*$ and $\bar{B}^*\phi$ dominate the decay.
All of the $1^+$ states are likely to be observed in the $\bar{B}^*\eta$ or $\bar{B}_s^*\bar{K}$ mode. For the two $2^+$ states, the dominant rearrangement decay channels for the higher and lower states are $\bar{B}_s\bar{K}^*$ and $\bar{B}^*\phi$, respectively.

\subsection{The $cs\bar{s}\bar{s}$ and $bs\bar{s}\bar{s}$ systems}

\begin{table}[htbp]
	\caption{Mass spectra for the $cs\bar{s}\bar{s}$ and $bs\bar{s}\bar{s}$ states in units of MeV. The thresholds of  $D_s\eta$  and $\bar{B}_s\eta$  are used to establish the lower limits for the tetraquark masses in the former and latter cases, respectively.   The upper limits are obtained using the effective quark masses: $m_s=542.4$ MeV, $m_c=1724.1$ MeV, and $m_b=5054.4$ MeV. We adopt values from the mass splitting model (fourth column) in the discussions.}\label{spectrum-Qsss}
	\scriptsize
	\begin{tabular}{c|ccccccc}\hline
		\hline\multicolumn{6}{c}{$cs\bar{s}\bar{s}$ states} \\\hline
		$I(J^{P})$ & $\langle H_{CM} \rangle$ &Eigenvalue  &Mass&Lower limits&Upper limits\\\hline
		$0(2^{+})$ &$\left(\begin{array}{c}72.8\end{array}\right)$&$\left(\begin{array}{c}72.8\end{array}\right)$&$\left(\begin{array}{c}3123.3\end{array}\right)$&$\left(\begin{array}{c}2853.0\end{array}\right)$&$\left(\begin{array}{c}3424.1\end{array}\right)$\\
		$0(1^{+})$ &$\left(\begin{array}{ccc}-15.2&24.8&11.7\\24.8&20.3&-93.3\\11.7&-93.3&-17.1\end{array}\right)$&$\left(\begin{array}{c}98.0\\-9.2\\-100.8\end{array}\right)$&$\left(\begin{array}{c}3148.5\\3041.2\\2949.7\end{array}\right)$&$\left(\begin{array}{c}2878.2\\2771.0\\2679.4\end{array}\right)$&$\left(\begin{array}{c}3449.3\\3342.1\\3250.5\end{array}\right)$\\
		$0(0^{+})$ &$\left(\begin{array}{cc}-59.2&161.7\\161.7&43.2\end{array}\right)$&$\left(\begin{array}{c}161.6\\-177.6\end{array}\right)$&$\left(\begin{array}{c}3212.0\\2872.9\end{array}\right)$&$\left(\begin{array}{c}2941.8\\2602.6\end{array}\right)$&$\left(\begin{array}{c}3512.9\\3173.7\end{array}\right)$\\
		\hline\hline\multicolumn{6}{c}{$bs\bar{s}\bar{s}$ states} \\\hline
		$I(J^{P})$ & $\langle H_{CM} \rangle$ &Eigenvalue  &Mass&Lower limits&Upper limits\\\hline
		$0(2^{+})$ &$\left(\begin{array}{c}53.1\end{array}\right)$&$\left(\begin{array}{c}53.1\end{array}\right)$&$\left(\begin{array}{c}6443.7\end{array}\right)$&$\left(\begin{array}{c}6161.4\end{array}\right)$&$\left(\begin{array}{c}6734.7\end{array}\right)$\\
		$0(1^{+})$ &$\left(\begin{array}{ccc}-11.5&60.0&28.3\\60.0&24.3&-68.4\\28.3&-68.4&6.9\end{array}\right)$&$\left(\begin{array}{c}93.3\\24.6\\-98.1\end{array}\right)$&$\left(\begin{array}{c}6484.0\\6415.2\\6292.5\end{array}\right)$&$\left(\begin{array}{c}6201.6\\6132.9\\6010.2\end{array}\right)$&$\left(\begin{array}{c}6774.9\\6706.2\\6583.5\end{array}\right)$\\
		$0(0^{+})$ &$\left(\begin{array}{cc}-43.7&118.6\\118.6&31.2\end{array}\right)$&$\left(\begin{array}{c}118.1\\-130.6\end{array}\right)$&$\left(\begin{array}{c}6508.7\\6260.1\end{array}\right)$&$\left(\begin{array}{c}6226.4\\5977.7\end{array}\right)$&$\left(\begin{array}{c}6799.7\\6551.0\end{array}\right)$\\
		\hline

		\hline
	\end{tabular}
\end{table}

 \begin{table}[htbp]
 	\caption{Rearrangement decays for the $cs\bar{s}\bar{s}$ and $bs\bar{s}\bar{s}$ cases. The two numbers in the parentheses for a decay channel mean dimensionless $100|\mathcal{M}|^2/{\mathcal C}^2$ and dimensional partial width, respectively. The masses and widths are presented in units of MeV.}\label{decay-Qsss}
 	\begin{tabular}{c|c|ccccc}\hline\hline
$I(J^{P})$&Mass&\multicolumn{4}{c}{Channels}&$\Gamma$\\	\hline
		\multicolumn{7}{c}{$cs\bar{s}\bar{s}$} \\\hline
 		&&$D_s^*\phi$&&&\\
 		$0(2^{+})$&$\left[\begin{array}{c}3123.3\end{array}\right]$&$\left[\begin{array}{c}(33.3, - )\end{array}\right]$&&&&$\left[\begin{array}{c}-\end{array}\right]$\\
 		&&$D_s^*\phi$&$D_s^*\eta$&$D_s^*\eta^{\prime}$&$D_s\phi$&\\
 		$0(1^{+})$&$\left[\begin{array}{c}3148.5\\3041.2\\2949.7 \end{array}\right]$&$\left[\begin{array}{c}(49.3, 111.2)\\(0.4, -)\\(0.4,-) \end{array}\right]$&$\left[\begin{array}{c}(0.4, 4.6)\\(3.4, 34.0)\\(15.8, 143.9) \end{array}\right]$&$\left[\begin{array}{c}(0.5, 2.3)\\(3.8, -)\\(17.7,-) \end{array}\right]$&$\left[\begin{array}{c}(3.3, 23.2)\\(26.1, 111.4)\\(12.3,-) \end{array}\right]$&$\left[\begin{array}{c}141.3\\145.5 \\143.9 \end{array}\right]$\\
 		&&$D_s^*\phi$&$D_s\eta$&$D_s\eta^{\prime}$&&\\
 		$0(0^{+})$&$\left[\begin{array}{c}3212.0\\2872.9\end{array}\right]$&$\left[\begin{array}{c}(54.9, 262.2)\\(3.4,  -)\end{array}\right]$&$\left[\begin{array}{c}(0.1, 0.7)\\(19.6, 209.6)\end{array}\right]$&$\left[\begin{array}{c}(0.1, 0.5)\\(22.0, -)\end{array}\right]$&&$\left[\begin{array}{c}263.4\\209.6 \end{array}\right]$\\
 		\hline
 		\multicolumn{7}{c}{$bs\bar{s}\bar{s}$} \\\hline
 		&&$\bar{B}_s^*\phi$&&&\\
 		$0(2^{+})$&$\left[\begin{array}{c}6443.7\end{array}\right]$&$\left[\begin{array}{c}(33.3, 14.5)\end{array}\right]$&&&&$\left[\begin{array}{c}14.5\end{array}\right]$\\
 		&&$\bar{B}_s^*\phi$&$\bar{B}_s^*\eta$&$\bar{B}_s^*\eta^{\prime}$&$\bar{B}_s\phi$&\\
 		$0(1^{+})$&$\left[\begin{array}{c}6484.0\\6415.2\\6292.5 \end{array}\right]$&$\left[\begin{array}{c}(45.1, 46.1)\\(3.4, -)\\(1.4, -) \end{array}\right]$&$\left[\begin{array}{c}(0.1, 0.3)\\(0.2, 0.4)\\(19.4, 45.7) \end{array}\right]$&$\left[\begin{array}{c}(0.1, 0.2)\\(0.2, 0.2)\\(21.8, -) \end{array}\right]$&$\left[\begin{array}{c}(8.9, 12.9)\\(30.2, 24.1)\\(2.5, -) \end{array}\right]$&$\left[\begin{array}{c}59.5\\24.7 \\45.7  \end{array}\right]$\\
 		&&$\bar{B}_s^*\phi$&$\bar{B}_s\eta$&$\bar{B}_s\eta^{\prime}$&&\\
 		$0(0^{+})$&$\left[\begin{array}{c}6508.7\\6260.1\end{array}\right]$&$\left[\begin{array}{c}(54.9, 68.6)\\(3.4,  -)\end{array}\right]$&$\left[\begin{array}{c}(0.1, 0.2)\\(19.6, 47.9)\end{array}\right]$&$\left[\begin{array}{c}(0.1, 0.1)\\(22.0, -)\end{array}\right]$&&$\left[\begin{array}{c}68.8\\47.9 \end{array}\right]$\\
 		\hline\hline
 	\end{tabular}
 \end{table}
 
The $Qs\bar{s}\bar{s}$ and the aforementioned $I=0$ $Qn\bar{s}\bar{n}$ tetraquarks resemble the excited $c\bar{s}$ mesons. At present, we do not consider the possible mixing among them. Similar to the $I=3/2$ $cn\bar{n}\bar{n}$ case where the Pauli principle matters, both the $cs\bar{s}\bar{s}$ and $bs\bar{s}\bar{s}$ systems have six tetraquark states. The masses and decay properties of the twelve states are given in Tables \ref{spectrum-Qsss} and \ref{decay-Qsss}, respectively. We illustrate the relative positions of the involved states in Fig. \ref{relative}(f) and Fig. \ref{relative}(l).
 
For the $cs\bar{s}\bar{s}$ case, the predicted tetraquarks have masses in the range of 2873$\sim$3212 MeV. The $J^P=2^+$ state with mass 3123.3 MeV is around the $D_s^*\phi$ threshold and some structure might be existent here. It may be searched for in the $D_s^+\gamma\phi$ channel in future experiments. Of the two $J^P=0^+$ states $\tilde{T}(2873)$ and $\tilde{T}(3212)$, the lower one has only one rearrangement decay channel $D_s\eta$, while the higher one mainly decays into $D_s^*\phi$ because of the stronger coupling. The three remaining $J^P=1^+$ states can all decay into $D_s^*\eta$, but each has a dominant rearrangement decay channel. The corresponding channels for the states, in order of increasing mass, are $D_s^*\eta$, $D_s\phi$, and $D_s^*\phi$. The lightest $\tilde{T}(2950)$ has only one decay channel. The heavier $\tilde{T}(3041)$ can additionally decay into $D_s^*\eta$ with a partial width ratio of $\Gamma(D^*_s\eta):\Gamma(D_s\phi)\sim0.3$. As for the heaviest $\tilde{T}(3149)$, the decay into $D_s\phi$ is also significant, with a partial width ratio of $\Gamma(D^*_s\phi):\Gamma(D_s\phi)\sim4.8$. The decay channels $D_s^*\eta$ and $D_s^*\eta^{\prime}$ for the $\tilde{T}(3149)$ are suppressed due to weak coupling.\\

The $bs\bar{s}\bar{s}$ case and the $cs\bar{s}\bar{s}$ case have the same color-spin wave functions. Their difference lies in the effective coupling strengths and phase spaces of decays. From Figs. \ref{relative}(j) and \ref{relative}(l), the two cases have similar features in spectrum. From the decay results listed in Table \ref{decay-Qsss}, the dominant decay channels for the lower and higher $J^P=0^+$ states are $\bar{B}_s\eta$  and $\bar{B}_s^*\phi$, respectively. For the three $1^+$ states $\tilde{T}(6293)$, $\tilde{T}(6415)$, and $\tilde{T}(6484)$, their dominant rearrangement decay channels are $\bar{B}_s^*\eta$, $\bar{B}_s\phi$, and $\bar{B}_s^*\phi$, respectively. The decay channel $\bar{B}_s^*\eta^\prime$ is open for $\tilde{T}(6415)$, while the $D_s^*\eta^\prime$ channel is not for the corresponding charmed state. The $2^+$ state is around the $\bar{B}_s^*\phi$ threshold. Searching in the $\bar{B}_s\phi\gamma$ invariant mass distribution may reveal some exotic structure.

\section{ SUMMARY}\label{secIV}

In this work, we study the mass spectra of the singly heavy tetraquark states by assuming that the $X(4140)$ is the lowest compact $cs\bar{c}\bar{s}$ tetraquark with $J^{PC}=1^{++}$. The spectra of tetraquarks are obtained by calculating the mass distances with respect to the reference $X(4140)$. The rearrangement decays are analyzed with a simple decay model. From numerical results, one finds that:
  
 (i) The recently observed state $T^a_{c\bar{s}0}(2900)^{++/0}$ by LHCb could be assigned as the second highest $I(J^P)=1(0^+)$ $cn\bar{s}\bar{n}$ state.

 (ii) The  $T_{cs0}^*(2900)^0$  is a good candidate of $I(J^P)=0(0^+)$ $cs\bar{n}\bar{n}$ tetraquark state.
 
 (iii) The lowest $bn\bar{s}\bar{n}$ tetraquark with $I(J^P)=1(0^+)$ is about 70 MeV above the $X(5568)$ reported by the $D0$ Collaboration. However, whether these two states are indeed related requires more experimental data.
  
 (iv) There are several possible stable singly heavy tetraquark states: the lightest $I(J^P)=0(0^+)$ $cn\bar{s}\bar{n}$ at around 2.2 GeV, the lightest $0(1^+)$ $cn\bar{s}\bar{n}$ around 2.4 GeV, and the lightest $0(0^+)$ $cs\bar{n}\bar{n}$ around 2.3 GeV. The corresponding bottom states are at around 5.6 GeV, 5.7 GeV, and 5.8 GeV, respectively. The $I(J^P)=0(1^+)$ $D\pi\bar{K}$ and $\bar{B}\bar{K}\gamma$ resonances near the $D^*\bar{K}$ and $\bar{B}^*\bar{K}$ thresholds, respectively, are likely narrow and warrant experimental searches.

If the observed singly charmed states $T^a_{c\bar{s}0}(2900)$ and $T_{cs0}^*(2900)^0$  are indeed tetraquark resonances, their partners should also exist. The rearrangement decay channels and partial width ratios we obtain should be helpful in studying such tetraquark states. Further experimental investigations are called for to validate these predictions.

\section{ACKNOWLEDGMENTS}

We thank all the members of our particle theory group for discussions and collaborations. This project was supported by the National Natural Science Foundation of China (No. 12235008, No. 12475143) and the Shandong Province Natural Science Foundation (No. ZR2023MA041).\\

\end{document}